\documentclass[12pt]{article}
\usepackage{changepage}

\usepackage[svgnames]{xcolor}
\usepackage[margin = 1in]{geometry}
\usepackage[T1]{fontenc}
\usepackage[utf8]{inputenc}
\usepackage[english]{babel}
 \usepackage{indentfirst}
 
\usepackage[normalem]{ulem}
\usepackage{amsfonts}
\usepackage{amssymb}
\usepackage{cancel}
\usepackage{amsmath}

\usepackage{enumerate}
\usepackage{bbm}
\usepackage{mathtools}
\usepackage{caption}
\usepackage{subcaption}
\usepackage{booktabs}	
\usepackage[nodisplayskipstretch]{setspace}
\usepackage{threeparttable}
\allowdisplaybreaks
\usepackage{tabularx}

\usepackage{graphicx}
\usepackage{fullpage}

\usepackage{xspace}
\usepackage{csquotes}

\usepackage{placeins}
\usepackage[footfont=normalsize,font=normalsize]{floatrow}
\usepackage{lscape}
\usepackage{longtable}
\usepackage{mathabx}
\usepackage{accents}
\usepackage{float}
\usepackage{enumitem}  
\usepackage{tikz}
\usetikzlibrary{calc,intersections}
  
	\usepackage[longnamesfirst]{natbib}
	\usepackage{bibunits}
	\defaultbibliography{Bibliography.bib}  
	\defaultbibliographystyle{aer}

\usepackage{hyperref}
\usepackage{url}
\hypersetup{
	colorlinks=true,
	linkcolor=Red,
	filecolor=magenta,      
	urlcolor=blue,
	citecolor=black
}
\usepackage{cleveref}

\usepackage{tcolorbox}
\newtcbox{\feedback}{nobeforeafter,colframe=black,colback=white,boxrule=0.5pt,arc=2pt,
	boxsep=0pt,left=2pt,right=2pt,top=2pt,bottom=2pt,tcbox raise base}

\usepackage{array}
\newcolumntype{L}[1]{>{\raggedright\let\newline\\\arraybackslash}m{#1}}
\newcolumntype{C}[1]{>{\centering\let\newline\\\arraybackslash\hspace{0pt}}m{#1}}
\newcolumntype{R}[1]{>{\raggedleft\let\newline\\\arraybackslash\hspace{0pt}}m{#1}}

\usepackage{ragged2e}
\newlength\ubwidth

\usepackage[bottom]{footmisc}

\newcommand\Item[1][]{%
	\ifx\relax#1\relax  \item \else \item[#1] \fi
	\abovedisplayskip=0pt\abovedisplayshortskip=0pt~\vspace*{-\baselineskip}}

\usepackage[hyperref]{ntheorem}
\theoremstyle{plain}
\theorembodyfont{\normalfont}

\newtheorem{thm}{Theorem}

\newtheorem{lem}{Lemma}

\newtheorem{proposition}{Proposition}

\newtheorem{asm}{Assumption}

\newtheorem{rem}{Remark}

\usepackage{adjustbox}

\usepackage{scrextend}

\usepackage{relsize}
\usepackage{multicol}
\usepackage{chngcntr}
\usepackage{etoolbox}
\usepackage{multirow}
\usepackage[labelfont=bf]{caption}

\usepackage{afterpage}
\usepackage{framed}
\usepackage{bigints}
\usepackage[titletoc,title, toc,page]{appendix}
\usepackage{xr}
\usepackage{hyphenat}

\deffootnote{1em}{1.6em}{\thefootnotemark \enskip}

\definecolor{lightgray}{gray}{0.9}
\numberwithin{equation}{section}
\numberwithin{thm}{section}
\numberwithin{Rem}{section}
\numberwithin{lemma}{section}
\numberwithin{lem}{section}
\numberwithin{algorithm}{section}
\numberwithin{rem}{section}
\numberwithin{proposition}{section}
\theorembodyfont{\normalfont}

\usepackage{silence}
\newcommand{\sE}{\mathop{}\!\mathbb{E}}
\newcommand{\sG}{\mathop{}\!\mathbb{G}}

\newcommand{\Fcal}{\mathcal{F}}

\newcommand{\bracks}[1]{\left[#1\right]}
\newcommand{\cbracks}[1]{\left\{#1 \right\}}

\newcommand{\hnull}{H_{0, P}}
\newcommand{\halt}{H_{1, P}}
\newcommand{\Pb}{\mathbf{P}}
\newcommand{\Pcal}{\mathcal{P}}
\newcommand{\varp}[1]{\mathbb{V}\text{ar}_P\bracks{#1}}

\newcommand{\var}[1]{\mathbb{V}\text{ar}\bracks{#1}}
\newcommand{\parens}[1]{\left(#1\right)}

\newcommand{\prob}[1]{\mathbb{P}\parens{#1}}

\newcommand{\Op}{\mathop{}\! O_{p}}
\newcommand{\Opnsq}{\mathop{}\! O_{p}\parens{n^{-1/2}}}
\newcommand{\op}{\mathop{}\! o_{p}}
\newcommand{\opnsq}{\mathop{}\! o_{p}\parens{n^{-1/2}}}
\newcommand{\oh}{\mathop{}\! o}

\newcommand{\convd}{\stackrel{d}{\rightarrow}}
\newcommand{\convp}{\stackrel{p}{\rightarrow}}

\newcommand{\R}{\mathbb{R}}
\DeclareMathOperator*{\argmin}{arg\,min}
\DeclareMathOperator*{\argmax}{arg\,max}

\newcommand{\ep}{e_{N_{p}, 1}}
\newcommand{\eq}{e_{N_{q}, 1}}

\newcommand{\Xcal}{\mathcal{X}}  
\newcommand{\Ycal}{\mathcal{Y}}  
	
\newcommand{\hhat}{ \widehat{h} }
\newcommand{\bhat}{ \widehat{b} }

\newcommand{\betahat}{\ensuremath{\widehat{\beta}}}

\newcommand{\dtall}{(d,t) \in \mathcal{S}}

\newcommand{\dtnpt}{(d,t) \in \mathcal{S}_{-} }
\newcommand{\dtnptp}{(d',t') \in \mathcal{S}_{-} }

\newcommand{\Jcal}{\mathcal{J}}

\newcommand{\Qx}[1]{\mathbf{Q}_{#1}(x_{c})}

\newcommand{\bZ}{\ubar{\mathbf{X}}}

\renewcommand{\u}{\mathbf{u}}

\newcommand{\Ifrac}{\mathfrak{I}}
\newcommand{\Ij}{\Ifrac_j}
\newcommand{\Imj}{\Ifrac_{-j}}
\newcommand{\sumj}{\sum_{j = 1}^J}
\newcommand{\sumIj}{\sum_{i \in \Ij}}

\newcommand{\sEj}{\sE_{n,j}}

\newcommand{\Jset}{\{1,...,J\}}

\newcommand{\dtamj}{(W_i)_{i \in \Imj}}

\newcommand{\rseq}{r_{n,d,t}}
\newcommand{\rseqo}{r_{n,1,1}}
\newcommand{\sseq}{s_{n,d,t}}
\newcommand{\supetap}{\sup_{\ptilde \in \Jcal_n^{p}}}
\newcommand{\supetam}{\sup_{\mtilde \in \Jcal_n^{m}}}
\newcommand{\tauyx}{\tau_{dr}(Y, X)}

\renewcommand{\k}{\mathbf{k}}

\newcommand{\mtilde}{\tilde{m}}

\newcommand{\ptilde}{\tilde{p}}
\newcommand{\phat}{\widehat{p}}

\newcommand{\mhat}{\widehat{m}}
\newcommand{\dti}{I_{d, t, i}}

\newcommand{\dt}{I_{d, t}}
\newcommand{\idx}[1]{\mathbbm{1}\{ #1 \}}

\newcommand{\xcdim}{\upsilon_{c}}

\newcommand{\xudim}{\upsilon_{u}}
\newcommand{\xodim}{\upsilon_{o}}

\newcommand{\xc}{x_{c}}
\newcommand{\xd}{x_{d}}
\newcommand{\xo}{x_{o}}
\newcommand{\xu}{x_{u}}

\newcommand{\kerps}{\widetilde{K}_{ps}}
\newcommand{\keror}{\widetilde{K}_{or}}

\newcommand{\Tcal}{\mathcal{T}}
\newcommand{\Vhat}{\widehat{V}_{n}}
\newcommand{\kec}{\mathbf{k!}}
\newcommand{\kabs}{\left | \mathbf{k} \right |}

\newcommand{\normnot}[2]{\mathcal{N}\parens{#1,\,#2}}

\newcommand{\chisq}[1]{\chi^2_{#1}}

\newcommand{\neigb}[1]{\mathcal{B}\parens{#1}}

\newcommand{\transpose}{\prime}

\newcommand{\deriv}[3]{#1^{\parens{#2}}_{#3}}

\newcommand{\hp}{h}

\newcommand{\php}{h}

\newcommand{\phor}{b_{d,t}}

\newcommand{\bhp}{    \php^{p + 1}  }

\newcommand{\bhor}{    \phor^{q + 1}  }

\newcommand{\sumi}{\dfrac{1}{n}\sum_{i = 1}^{n}}

\newcommand{\what}{\widehat{w}}

\newcommand{\gammahat}{\widehat{\gamma}}

\newcommand{\Omegahat}{\widehat{\Omega}}
\newcommand{\etahat}{\widehat{\eta}}

\newcommand{\tauhat}{\widehat{\tau}}

\newcommand{\ubar}[1]{\underaccent{\bar}{#1}}

\def\ubar#1{\underline{#1}}

\newcommand{\normabs}[1]{\left\lvert#1\right\rvert}
\newcommand{\norm}[1]{\left\lVert#1\right\rVert}

\newcommand{\norms}[1]{  \left\lVert#1\right\rVert_{\infty}}
\newcommand{\normL}[1]{  \left\lVert#1\right\rVert_{L_{2}}}

\makeatletter
\newcommand*{\indep}{%
  \mathbin{%
    \mathpalette{\@indep}{}%
  }%
}
\newcommand*{\nindep}{%
  \mathbin{
    \mathpalette{\@indep}{/}%
  }%
}
\newcommand*{\@indep}[2]{%
  \sbox0{$#1\perp\m@th$}
  \sbox2{$#1=$}
  \sbox4{$#1\vcenter{}$}
  \rlap{\copy0}
  \dimen@=\dimexpr\ht2-\ht4-.2pt\relax
  \kern\dimen@
  \ifx\\#2\\%
  \else
    \hbox to \wd2{\hss$#1#2\m@th$\hss}%
    \kern-\wd2 %
  \fi
  \kern\dimen@
  \copy0 
}
\makeatother

\newcommand\independent{\protect\mathpalette{\protect\independentT}{\perp}}
\def\independentT#1#2{\mathrel{\rlap{$#1#2$}\mkern2mu{#1#2}}}
\makeatletter
\newcommand*{\addFileDependency}[1]{
  \typeout{(#1)}
  \@addtofilelist{#1}
  \IfFileExists{#1}{}{\typeout{No file #1.}}
}
\makeatother

\newcommand*{\myexternaldocument}[1]{%
    \externaldocument{#1}
    \addFileDependency{#1.tex}
    \addFileDependency{#1.aux}
}

\myexternaldocument{supp}
\begin{document}
\begin{bibunit}

\title{Difference-in-Differences with Compositional Changes}  
\author{Pedro H. C. Sant'Anna\thanks{Emory University.
		E-mail: pedro.santanna@emory.edu}
	\and Qi Xu\thanks{Department of Health Outcomes and Biomedical Informatics, University of Florida. E-mail:
		qixu@ufl.edu}}
\date{\today}

\maketitle
\thispagestyle{empty}
\begin{abstract}
    
This paper studies Difference-in-Differences (DiD) setups with repeated cross-sectional data and potential compositional changes across time periods. We begin our analysis by deriving the efficient influence function and the semiparametric efficiency bound for the average treatment effect on the treated (ATT). We introduce nonparametric estimators that attain the semiparametric efficiency bound under mild rate conditions on the estimators of the nuisance functions, exhibiting a type of rate doubly robust (DR) property. Additionally, we document a trade-off related to compositional changes: We derive the asymptotic bias of DR DiD estimators that erroneously exclude compositional changes and the efficiency loss when one fails to correctly rule out compositional changes. We propose a nonparametric Hausman-type test for compositional changes based on these trade-offs. The finite sample performance of the proposed DiD tools is evaluated through Monte Carlo experiments and an empirical application. We consider extensions of our framework that accommodate double machine learning procedures with cross-fitting, and setups when some units are observed in both pre- and post-treatment periods. As a by-product of our analysis, we present a new uniform stochastic expansion of the local polynomial multinomial logit estimator, which may be of independent interest.

\end{abstract}

\newpage

\setcounter{page}{1}

\section{Introduction} \label{sec::introduction}
Difference-in-Differences (DiD) designs have been used widely for identifying and estimating causal effects with observational data. Identification in this research design typically relies on a conditional parallel trends assumption stipulating that conditional on a set of covariates, the average untreated outcomes among treated and comparison groups would have evolved ``in parallel''.  When one pairs this assumption with common support and no-anticipation assumptions, it is easy to establish that the average treatment effect on the treated (ATT) is nonparametrically identified when panel data is available. When one only observes repeated cross-sectional data, it is common to impose further a no-compositional change assumption, also known as the stationarity assumption. This is the case in the widely cited DiD procedures of \citet{Heckman1997}, \citet{Abadie2005}, \citet{Sant2020doubly}, and \citet{callaway2021difference}, for example.

Although we have seen a lot of recent developments in DiD methods (see \citealp{Roth2023} for an overview of recent DiD developments), little attention has been paid to understanding the importance and limitations of the no-compositional changes assumption. This paper aims to fill this gap by providing researchers with new tools that can be used when they doubt such an assumption and/or to test its plausibility.

Before discussing the paper's contributions, it is worth stressing why ruling out compositional changes across periods can be restrictive in real empirical applications. Essentially, the no-compositional changes assumption requires one to sample observations from the same population across periods, which can be unrealistic in some scenarios. For example, \citet{Hong2013} studies the effect of Napster on recorded music sales.  He uses data from the 1996–2002 Interview Surveys of the Consumer Expenditure Survey.  Over this period,  the composition of internet users has changed substantially.  The early adopters tend to be younger, richer, more educated, and technically savvy, whereas later adopters exhibit a higher diversity level in demographics. If one ignores such imbalances of group composition across time, the (negative) effect of Napster on music sales can be overestimated, as the decrease in the average music expenditure may be attributed to a post-Napster group with more households having low reservation prices for recorded music. Other applications also share this concern, as discussed below and in more detail in Section \ref{sec::application}. Therefore, having causal inference tools that can assess if the findings are robust to compositional changes in the sample is of practical interest.

We begin our analysis by showing that one can identify the ATT in DiD setups without invoking the no-compositional changes assumption. In this scenario, we derive the efficient influence function (EIF) and the semiparametric efficiency bound for the ATT. We then form generic nonparametric estimators built on the EIF that can achieve the semiparametric efficient bound under mild smoothness conditions, a rate doubly robust (DR) property \citep{Smucler2019}. Heuristically, this rate DR property allows for a trade-off between the rate of convergence of the two nuisance estimators. It implies that nonparametric DiD estimators for the ATT based on the EIF are $\sqrt{n}$-consistent and asymptotically normal even if one of the outcome regression or generalized propensity score functions is very complex so long as the other is simple enough; this is weaker then requiring that the estimators for \textit{both} nuisance functions converge sufficiently fast. These results are general and do not rely on a specific choice of estimators for nuisance functions. Nonetheless, they do not help us with practical inference procedures. For that, we use a local polynomial estimator for the outcome-regressions models and the local multinomial logit regression to estimate the generalized propensity score, the latter of which is fairly new in the DiD literature. Importantly, our nonparametric estimators can accommodate both discrete and continuous covariates,\footnote{As a side contribution of this paper, we provide a new result on the uniform expansion of the local (multinomial) logit estimators, which accommodates both continuous and discrete variables. This result may be of independent interest.} and all tuning parameters are selected in a data-driven way via cross-validation.\footnote{Bandwidths are selected independently for each set of nuisance functions instead of directly for the ATT estimator. Nonetheless, the second-step asymptotic results remain unaffected, as the double-robust rate property of our proposed estimator ensures that using optimal bandwidths for each nuisance function still enables valid inference on the ATT estimator.}  Finally, we show that the estimtor proposed by \citet{Sant2020doubly} is no longer DR in this DiD setup with compositional changes. In fact, we show that even when all nuisance functions are correctly specified, the \citet{Sant2020doubly}'s DR DiD estimand does not identify the ATT in this general setup. Overall, this first set of results highlights ``the best'' one can do in DiD setups with compositional changes.

Next, we tackle the problem of how much efficiency one may lose by not exploring the no-compositional change assumption when it is valid. To answer this question, we compare our derived semiparametric efficiency bound that does not impose the no-compositional changes assumption with the semiparametric efficiency bound derived by \citet{Sant2020doubly} that fully exploits it. As expected, the extra layer of robustness comes at the cost of loss of efficiency. Heuristically speaking, the no-compositional change assumption allows one to pool the covariate data from all time periods, substantially increasing the effective sample size and the precision of the DiD estimator compared to the one that does not impose the no-compositional change assumption. Regarding the estimation of nuisance functions, we also note that, under the no-compositional changes assumption, one can use standard (binary) propensity score estimators. However, when one allows for compositional changes, one needs to use generalized propensity score estimators, as there are now four effective groups depending on the treatment group and the time the unit is observed.

In practice, determining whether compositional changes are a concern for a given empirical application is not always obvious. Specifically, it is unclear whether imposing a no-compositional change assumption will lead to biased ATT estimates. Using our previous results, we propose a nonparametric \citet{hausman1978specification}-type test for no-compositional changes. The test compares our nonparametric DiD estimator of the ATT, which is robust to compositional changes, with the nonparametric extension of \citet{Sant2020doubly}'s DR DiD estimator, which assumes no compositional changes. We derive the large sample properties of the proposed test, which shows that it controls size asymptotically and is consistent against a broad set of alternatives.

We demonstrate the practical appeal of our proposed DiD tools through Monte Carlo simulations and an empirical application that revisits \citet{sequeira2016corruption}. She leverages a quasi-experimental variation created by a large reduction in the average nominal tariff rate between South Africa and Mozambique in 2008 to study the causal effect of tariff rate reduction on trade costs and corruption behavior using a two-way fixed effects specification with covariates that implicitly imposes a no-compositional changes assumption, among other arguably unnecessary homogeneity assumptions. We use our nonparametric tests to assess the plausibility of the no-compositional changes assumption and fail to reject it at the usual significance levels. Our results support the conclusions by \citet{sequeira2016corruption} that tariff liberalization decreases corruption, and our DR DiD estimates are similar to those in the original paper.

Finally, we consider some extensions of our framework. Although our paper primarily focuses on leave-one-out estimators for the nuisance functions, we also consider cross-fitted estimators, drawing on the modern double machine learning literature; see, e.g., \citet{Farrell2015}, \citet{Belloni2017}, \citet{Chernozhukov2017}, \citet{Colangelo_Lee_2023}, and \citet{Kennedy_2023_DR_review}. An important difference between cross-fitted and leave-one-out estimators is that the former requires the number of folds $J$ to be fixed as $ n \to \infty$, while the latter allows it to grow with sample size (as $J=n$). As a result, these estimation procedures rely on different types of assumptions and require different proof strategies to establish their large-sample properties.

We also extend our analysis to applications where some units are observed in both pre- and post-treatment periods (a balanced panel data component), while the remaining units are observed in only one of the two periods (a cross-sectional component). However, our main results focus on sampling schemes without overlapping units, thus precluding these scenarios. To cover this practically relevant setup, we derive the EIF and the semiparametric efficiency bound for the ATT in setups that allow for this type of sampling scheme. Similar to the pure repeated cross-sectional setup, we discuss how one can build on the derived EIF to form nonparametric estimators that achieve the semiparametric efficiency bound.

\textbf{Related literature:} This article belongs to the extensive literature on semiparametric DiD methods.  We refer the reader to \citet{Roth2023} and \citet{Baker_etal_DiD_practioner_guide_2025} for a synthesis of recent advances in DiD. Within this broad literature, the paper closest to ours is \citet{Sant2020doubly}, which proposes DR DiD estimators for the ATT and derives the semiparametric efficiency bound for such estimators, too. In sharp contrast to us, though, all the results in \citet{Sant2020doubly} rely on a no-compositional change assumption. Thus, our results complement theirs. Furthermore, \citet{Sant2020doubly}'s theoretical results rely on parametric first-step estimators, while we accommodate nonparametric estimators. A perhaps side and minor contribution of our paper is establishing the statistical properties of \citet{Sant2020doubly}'s DR DiD estimator with nonparametric estimates of the nuisance functions; see also \citet{chang2020double}.

Our paper also relates to the causal inference literature on compositional changes over time. \citet{Hong2013} develops a matching-based estimator under a ``selection-on-observable''-type assumption, which is different and arguably stronger than our conditional parallel trends assumption. \citet{Hong2013} also does not discuss efficiency issues as we do. \citet{Stuart2014} propose inverse probability weighted estimators for the ATT in DiD setups under compositional changes. In contrast to us, their estimator does not enjoy any DR property and may not attain the semiparametric efficiency bound. \citet{nie2019nonparametric} is also interested in DiD estimators under compositional changes. Their estimator substantially differs from ours: they use meta-learners and cross-fitting to estimate nuisance functions, while our estimator is based on the EIF for the ATT. When treatment effects are heterogeneous, their estimators do not target the ATT but the ATE, which, in our context, is not identified. They do not consider tests for the no-compositional changes assumption as we do. 

Finally, we contribute to the semiparametric two-stage estimation that depends on nonparametrically estimated functions.  See, e.g., \citet{Newey1994}, \citet{Chen2003}, \citet{Chen2008}, \citet{Ackerberg2014}, \citet{rothe2019properties}, among many others. Our results on local multinomial logit regression builds on \citet{fan1995local}, \cite{claeskens2003bootstrap}, \cite{li2005uniform}, and \citet{Kong2010}. The novel result on the uniform expansion of the local multinomial logit estimator may be of independent interest. Our cross-fitting double machine learning results build on, among others, \citet{Belloni2017}, \citet{Chernozhukov2017}, \citet{Colangelo_Lee_2023}, and \citet{Kennedy_2023_DR_review} and extend their cross-sectional results under unconfoundedness to the DiD context.  

\textbf{Organization of the paper:} Section \ref{sec::identification} introduces the identification framework of the DiD parameter under compositional changes, presents the semiparametric efficiency results, and discusses the bias-variance trade-off of ruling out compositional changes. In Section \ref{sec::estimation}, we present our nonparametric DR DiD estimators, discuss their large sample properties, and how to pick tuning parameters. Section \ref{sec::test} discusses a test for no-compositional changes.  Monte Carlo simulations are provided in Section \ref{sec::simulation}, and an empirical illustration is considered in Section \ref{sec::application}. Section \ref{sec::extensions} discusses extensions. Proofs and additional results are reported in the Supplemental Appendix available \href{https://psantanna.com/files/DiD_CC_Appendix.pdf}{here}.

\section{Difference-in-Differences} \label{sec::identification}
\subsection{Framework} \label{sec:setup}
This section describes our setup. We focus on the canonical two-period and two-group setup for conciseness and transparency.  We have two time periods, $t=0$, where no unit is exposed to the treatment, and time $t=1$, where units in the group with $D=1$ are exposed to treatment; here, $D$ is a binary treatment indicator. We adopt the potential outcome notation where $Y_{it}\left( 0\right) $ and $Y_{it}\left( 1\right) $ denote the untreated and treated potential outcome for unit $i$ at time $t$, respectively. 
Observed outcomes are given by $Y_{it} = D_{it}Y_{it}(1) + (1 - D_{it})Y_{it}(0)$. We also assume that a $k$-dimensional vector of pre-treatment characteristics $X_{i}\in \mathcal{X} \subseteq \mathbb{R}^{k}$ is available.

This paper considers the case where one has access to repeated cross-sectional data. To formalize this idea, let $T_{i}$ be a dummy variable that takes value one if the observation $i$ is observed only in the post-treatment period $t=1$, and zero if observation $i$ is only observed in the pre-treatment period $ t=0 $. Define $Y_{i}=T_{i}Y_{i1}+\left( 1-T_{i}\right) Y_{i0}$, and let $ n_{1}$ and $n_{0}$ be the sample sizes of the post-treatment and pre-treatment periods such that $n=n_{1}+n_{0}$. 

\begin{asm}[Sampling]\label{ass::sampling}
The pooled data $ \left\{ Y_{i},D_{i},X_{i},T_{i}\right\} _{i=1}^{n}$ consists of independent and identically distributed draws from the mixture distribution $ P \in \mathcal{P}$: %
\begin{eqnarray*}
\prob{Y\leq y,D=d,X\leq x,T=t}&=&t\cdot \prob{T = 1} \cdot \prob{Y_{1}\leq y,D=d,X\leq x|T=1} \\
&&+\left( 1-t\right) \cdot \prob{T = 0} \prob{Y_{0}\leq y,D=d,X\leq x|T=0},
\end{eqnarray*}%
where $\left( y,d,x,t\right) \in \Ycal \times  \left\{ 0,1\right\} \times \Xcal \times \left\{ 0,1\right\} .$
\end{asm}

Assumption \ref{ass::sampling} allows for different sampling schemes. For instance, it accommodates the binomial sampling scheme where an observation $i$ is randomly drawn from either $\left( Y_{1},D,X\right) $ or $\left( Y_{0},D,X\right) $ with a fixed probability. It also accommodates the  ``conditional'' sampling scheme where $n_{1}$ observations are sampled from $\left( Y_{1},D,X\right) $, $n_{0}$ observations are sampled from $\left( Y_{0},D,X\right) $ and $\prob{T = 1} =n_{1}/n$. 
Importantly, Assumption \ref{ass::sampling} does not impose that we are sampling from the same underlying distribution across time periods, implying that it is fully compatible with compositional changes (\citealp{Hong2013}). This is in contrast to most of the DiD literature. For example, Assumption 1(b) in \citet{Sant2020doubly} explicitly imposes that $ (D, X) \independent T$; see also \citet{Heckman1997}, and \citet{Abadie2005} for other DiD procedures that rely on this stationarity condition.

As is typical in DiD setups, we are interested in the average treatment effect in time period $t=1$ among the treated units,
\begin{equation}\label{eq.att.def}
ATT= \tau = \sE\left[
Y_{1}\left( 1\right) |D=1,T=1\right] - \sE\left[ Y_{1}\left( 0\right) |D=1,T=1\right] .
\end{equation}%

Given that the untreated potential outcome $Y_{i1}(0)$ is never observed for the treated units, we need to impose assumptions to uncover $\sE\left[ Y_{1}\left( 0\right) |D=1, T=1\right]$  from the data. We make conditional parallel trends, no-anticipation, and strong overlap assumptions toward this goal. Let $\mathcal{S} \equiv \{0, 1\}^2$  and  $\mathcal{S}_{-} \equiv \{(1,0), (0,1), (0,0)\}$.  

\begin{asm}[Conditional parallel trends, no-anticipation, and overlap] \label{ass::ident} \phantom{a} \linebreak
For some $\varepsilon >0$, $ \dtnpt$, and for almost every $x \in \mathcal{X}$,
\begin{align*}
(i)& \quad\quad \sE[Y_{1}(0)|D=1,T=1,X=x] - \sE[Y_{0}(0)|D=1,T=0,X=x] \\
              &\quad\quad\quad\quad\quad \quad\quad\quad\quad\quad\quad\quad = \sE[Y_{1}(0)|D=0,T=1,X=x]  -\sE[Y_{0}(0)|D=0,T=0,X=x].\\
 (ii)& \quad \quad  \sE[Y_{0}(0)|D=1,T=0,X=x] = \sE[Y_{0}(1)|D=1,T=0,X=x].\\
 (iii)& \quad\quad \mathbb{P}\left( D=1,T=1\right) >\varepsilon \text{ and } \prob{D=d,T=t|X=x} \geq \varepsilon.
\end{align*}

\end{asm}
Assumption \ref{ass::ident}(i) is the conditional parallel trends assumption (CPT) stating that conditioning on $X$, the average evolution of the untreated potential outcome is the same among the treated and untreated groups. This assumption allows for covariate-specific trends and does not restrict the trends among different covariate strata. This assumption is commonly used by DiD methods with repeated cross-sectional data---see, e.g., \citet{Hong2013}, \citet{Stuart2014}, \citet{nie2019nonparametric}, \citet{Sant2020doubly}, \citet{callaway2021difference}, among others. When $Y_t(0)$ is mean independent of $T$ conditional on $D$ and $X$, Assumption \ref{ass::ident}(i) reduces to the CPT assumption commonly stated for balanced panel data---we take this to be the leading case, in which all usual tools to assess the plausibility of parallel trends assumptions can be used. In the presence of sample selection, though, additional challenges related to the source of missingness-not-at-random can arise, potentially leading to partial identification of the parameter of interest \citep{Rathnayake_etal_2024_DiD_SampleSelection}; we defer a full treatment of this topic to future research. 
    
Assumption \ref{ass::ident}(ii) is a no-anticipation assumption (NAA) stating that, on average, treated units do not act on the future treatment prior to its implementation \citep{Abbring2003, Malani2015}. Assumption \ref{ass::ident}(iii) is an overlap condition that guarantees that there are some treated units in the post-treatment period and that the covariates do not fully determine treatment status. This condition ensures nonparametric regular inference procedures \citep{Khan2010}.

\subsection{Identification and semiparametric efficiency bound}\label{sec::id_and_efficiency}
Under Assumptions \ref{ass::sampling} and \ref{ass::ident}, it is straightforward to show that the ATT is nonparametrically identified by the outcome regression estimand\footnote{See Lemma \ref{lem:id} in Appendix \ref{sec:prof.ident} for the formalization of these results.}
\begin{equation}
\tau =  \tau_{or} \equiv \sE\left[ \left.  Y \right\vert D=1,T=1 \right]  - \sE\left[ \left. m_{1,0}(X) + m_{0,1}(X) - m_{0,0}(X) \right\vert D=1,T=1 \right], \label{OR}
\end{equation}%
where $ m_{d,t}(x) = E[Y|D= d,T = t,X = x] $.  Alternatively, it is also easy to show that one can identify the ATT using an inverse probability weighted estimand
\begin{equation}
\tau = \tau_{ipw} \equiv \sE\left[ \left( w_{1,1}(D, T) - w_{1, 0}(D, T,X)  - w_{0, 1}(D, T, X) + w_{0, 0}(D, T,X )\right)
Y \right], \label{IPW}
\end{equation}
where, for $ \dtnpt $
\begin{align}
	w_{1,1}(D, T)  &=  \dfrac{DT }{\sE[DT]} , \nonumber \\
	w_{d, t}(D, T, X) & = \left. \dfrac{I_{d,t} \cdot p(1,1,X) }{p(d, t, X)}\middle / \sE\bracks{\dfrac{I_{d,t} \cdot p(1,1,X) }{p(d, t, X)}}, \right.  \label{eq.weights} 
\end{align}
$I_{d,t} = \idx{D = d, T = t}$, and $ p(d,t, x) = \prob{D = d, T = t | X = x} $ is a so-called generalized propensity score. Notice that the weights in \eqref{eq.weights} are of the \citet{Hajek1971}-type. This guarantees that all the weights sum up to one and typically results in more stable finite sample behavior; see, e.g., \citet{millimet2009specification, Busso2014, Sant2020doubly}.

From \eqref{OR} and \eqref{IPW}, it is clear that any linear combination of $\tau_{or}$ and $\tau_{ipw}$ also identifies the ATT under our assumptions. There are also many other potential estimands that make use of nonlinear combinations of the different terms in $\tau_{or}$ and $\tau_{ipw}$ and identify the ATT. From this simple observation, a natural question that arises is: How can we combine these two strategies to obtain an efficient estimator for the ATT? The next theorem addresses this question through the lens of semiparametric efficiency theory. Specifically, we derive the EIF for the ATT under Assumptions \ref{ass::sampling} and \ref{ass::ident}, as well as its semiparametric efficiency bound. This bound represents the maximum precision achievable in this context under the given assumptions. As so, it provides a benchmark that researchers can use to assess whether any given (regular) semiparametric DiD estimator for the ATT fully exploits the empirical content of Assumptions \ref{ass::sampling} and \ref{ass::ident}.\footnote{To simplify exposition, we abstract from additional technical discussions related to the conditions to guarantee quadratic mean differentiability and their implications for the precise definition of EIF; see, e.g., Chapter 3 of \cite{Bickel1998} for more details.} Hereafter, let $ \tauyx = Y - (m_{1,0}(X) + (m_{0,1}(X) - m_{0,0}(X)) )$ and $W = (Y,D,X, T)$. We also denote the ATT by $\tau$.
\begin{thm}(Semiparametric efficiency bound)\label{thm::seb}
Suppose Assumptions  \ref{ass::sampling} and  \ref{ass::ident} hold. Then, the EIF for $ \tau $ is given by
	\begin{equation}
\eta_{\text{eff}}(W) =  w_{1, 1}(D, T)(\tauyx -\tau) +  \sum_{\dtnpt}   (-1)^{(d + t)}  w_{d,t}(D, T, X)  (Y - m_{d,t}(X)),\label{eff_inf_funct}
	\end{equation}
where the weights are defined in \eqref{eq.weights}. Furthermore, the semiparametric efficiency bound for the set of all regular estimators of $ \tau $ is 
	\begin{equation*} 
	\sE[\eta_{\text{eff}}(W)^{2}] = \dfrac{1}{\sE\left[ DT\right]^{2}}\sE\left[ DT(\tauyx - \tau)^{2} 	+\sum_{\dtnpt}\dfrac{I_{d,t} \cdot p(1,1,X)^{2}}{p(d,t,X)^{2}} (Y - m_{d,t}(X))^{2}\right].
	\end{equation*}
\end{thm}
 
Apart from providing an efficiency benchmark, Theorem \ref{thm::seb} also provides us a template to construct efficient estimators for $\tau$. That is, given that any influence function has a mean of zero, we can take the expected value of $\eta_{\text{eff}}(W)$ and isolate $\tau$ to get the following estimand for the ATT
\begin{align}
\tau = \tau_{dr} &\equiv \sE\left[ w_{1, 1}(D, T)\tauyx +  \sum_{\dtnpt}   (-1)^{(d + t)}  w_{d,t}(D, T, X)  (Y - m_{d,t}(X))\right].\label{eq.dr}
\end{align}
We discuss in Section \ref{sec::estimation} how to leverage \eqref{eq.dr} to nonparametrically estimate $\tau$ under mild assumptions.



\subsection{Bias-Variance trade-off with respect to stationarity} \label{sec:bias.var.tradeoff}
All the estimands described in Section \ref{sec::id_and_efficiency} account for compositional changes over time. As discussed the introduction, most DiD estimators typically assume no compositional changes \textit{a priori}. A natural question then arises: How biased would these estimators be when they erroneously rule out compositional changes?

To tackle this question, we examine the bias of the semiparametrically efficient DiD estimator for the ATT proposed by \citet{Sant2020doubly} that rule out compositional changes. Before diving into this analysis, we need to introduce some additional notation and clarify the assumptions, estimands, and other aspects of \citet{Sant2020doubly}'s approach.

First, \citet{Sant2020doubly} explicitly rules out compositional changes by relying on the following stationarity assumption.

\begin{asm}[Stationarity]\label{ass::stationarity}
	$ (D, X) \independent T.$
\end{asm}

Intuitively, Assumption \ref{ass::stationarity} enables researchers to pool covariates and treatment variables from both periods. As a result, under Assumption \ref{ass::stationarity}, it follows that $\sE\left[D|X,T=1\right] =\sE\left[D|X\right]\equiv \tilde{p}(X)$, which also affects the definition of the ``relevant'' propensity score. Indeed, as discussed in \citet{Sant2020doubly}, one can identify the ATT under Assumption \ref{ass::stationarity} using the IPW estimand
\begin{align}
\tau_{ipw,ncc} &\equiv \sE\left[ \left(w^{sz}_{1,1}(D, T, X)  - w^{sz}_{1,0}(D, T, X) - w^{sz}_{0,1}(D, T, X) + w^{sz}_{0,0}(D, T, X)\right) Y \right],\label{eq.sz_ipw}
\end{align}
where, for $t=0,1$,
\begin{eqnarray}
w_{1,t}^{sz}\left( D,T, X\right) &=&\frac{D\cdot \idx{T=t} }{\mathbb{E}\left[ D\cdot \idx{T=t} \right] }, \nonumber \\
w_{0,t}^{sz}\left( D,T,X\right) &=&\left. \frac{\tilde{p}(X)\left( 1-D\right)
\cdot \idx{T=t} }{1-\tilde{p}(X)}\right/ \mathbb{E}\left[ \frac{\tilde{p}(X)\left( 1-D\right) \cdot \idx{T=t} }{1-\tilde{p}(X)}\right];\label{eq:sz_weights}
\end{eqnarray}
see also \citet{Abadie2005}. When one compares the IPW weights in \eqref{eq.sz_ipw} with the IPW weights in \eqref{IPW}, it is clear that $w_{1,1}^{sz}(\cdot) = w_{1,1}(\cdot)$ whereas the remaining three IPW weights differ. Under Assumption \ref{ass::stationarity}, one only needs to consider binary propensity score models, and use these simpler functions to construct IPW weights for untreated units (in both periods) in \eqref{eq:sz_weights}. When one allows for compositional changes such that Assumption \ref{ass::stationarity} is potentially violated, the distribution of the covariates may vary over time. Thus, one needs to consider generalized propensity scores, as now one has four groups depending on the treatment group and the period that a unit is observed. This affects the IPW weights in \eqref{eq.weights} not only for the untreated units, but also for the treated units observed only in the pre-treatment period. This discussion sheds light on how Assumption \ref{ass::stationarity} ``simplifies'' the construction of IPW estimators.

We note that Assumption \ref{ass::stationarity} also allows one to leverage more data than \eqref{OR} when constructing an outcome regression estimand. More precisely, under Assumption \ref{ass::stationarity}, one can identify the ATT using the following regression-based estimand\footnote{Under Assumption \ref{ass::stationarity}, one can also use the alternative regression-based estimand that leverages more data than $\tau_{or, ncc}$ defined as $\tau_{or, ncc,2} \equiv  \sE\left[ \left. m_{1,1}(X) - m_{1,0}(X) - m_{0,1}(X) + m_{0,0}(X) \right\vert D=1 \right]$. We do not discuss this further as we directly address estimators based on the EIF in \eqref{eq.sz}. }
\begin{equation}
\tau_{or, ncc} \equiv \sE\left[ \left.  Y \right\vert D=1,T=1 \right]  -  \sE\left[ \left.  m_{1,0}(X) + m_{0,1}(X) - m_{0,0}(X) \right\vert D=1 \right]; \label{OR_SZ}
\end{equation}%
see, e.g., \citet{Sant2020doubly}. To see the difference between \eqref{OR_SZ} and \eqref{OR}, note that the last term in \eqref{OR_SZ} integrates the covariates $X$ using the distribution of treated units from both periods, i.e., the pooled treated distribution. This is valid under Assumption \ref{ass::stationarity}. When one allows for compositional changes over time, as in \eqref{OR}, one needs to integrate over the covariate distribution among treated units observed in the post-treatment period only. 

Just like in Section \ref{sec::id_and_efficiency}, one can combine $\tau_{or, ncc}$ and $\tau_{ipw,ncc}$ to get more efficient estimators under Assumption \ref{ass::stationarity}. Indeed, \citet{Sant2020doubly} show that, under Assumptions \ref{ass::sampling}, \ref{ass::ident}, and \ref{ass::stationarity}, the EIF for the ATT is given by 
\begin{align}
 	\eta_{sz}(W) &=  \dfrac{D}{\sE[D]}\bigg( \tau(X) -\tau\bigg) + \sum_{\dtall}   (-1)^{(d + t)}  w^{sz}_{d,t}(D, T, X)  (Y - m_{d,t}(X)),\label{eq.SZ_score}
 \end{align}
where $ \tau(x) = (m_{1,1}(x) - m_{1,0}(x)) - (m_{0,1}(x) - m_{0,0}(x))$ is the conditional ATT (CATT). Based on \eqref{eq.SZ_score}, \citet{Sant2020doubly} propose the following DR estimand for the ATT,
\begin{align}
\tau_{sz} &\equiv \sE\left[\dfrac{D}{\sE[D]}\tau(X) + \sum_{\dtall}   (-1)^{(d + t)}  w^{sz}_{d,t}(D, T, X)  (Y - m_{d,t}(X))\right].\label{eq.sz}
\end{align}
Heuristically, $\tau_{sz}$ can be viewed as the analog of $\tau_{dr}$ when, on top of the identification assumptions, one assumes that there are no compositional changes. Compared to $\tau_{dr}$, it is based on binary propensity scores and integrates the conditional ATT using data from both time periods.

The following proposition shows that $\tau_{sz}$ does not recover the ATT when Assumption \ref{ass::stationarity} is potentially violated, i.e., under compositional changes. It also precisely quantifies the bias relative to $\tau_{sz}$.

\begin{proposition}\label{prop::bias} Under Assumptions \ref{ass::sampling} and \ref{ass::ident}, we have that
\begin{align*}
 \tau_{sz} - \tau_{dr} = & \sum_{\dtall}   (-1)^{(d + t)} \sE\left[\bigg(\dfrac{D}{\sE[D]} - \dfrac{DT}{\sE[DT]}\bigg) m_{d,t}(X) \right]\\
&  +\sum_{\dtnpt}   (-1)^{(d + t)} \sE\left[ \left( w^{sz}_{d,t}(D, T, X)  - w_{d,t}(D, T, X) \right)(Y - m_{d,t}(X))\right]\\
 = &\sE[ \tau(X) |D =1]  - \sE[ \tau(X) | D = 1, T = 1 ] \\
 = & \sE[ \tau(X) |D =1]  - \tau.
\end{align*}
\end{proposition}



Proposition \ref{prop::bias} provides bias decomposition for $ \tau_{sz} $ when the stationarity assumption is not imposed. The first equality in Proposition \ref{prop::bias} follows from a direct comparison between our proposed estimand for the ATT and the one proposed by \citet{Sant2020doubly}, while the second equality is a consequence of the law of iterated expectations.\footnote{Here, we are implicitly considering the case where there are no (global) model misspecifications, which aligns with the fully nonparametric approach we adopt. One can compute a similar bias decomposition when one adopts parametric working models for the nuisance functions, though the notation becomes much more cumbersome. } The third equality is due to the definition of ATT and Assumptions \ref{ass::sampling} and \ref{ass::ident}. These calculations show that \citet{Sant2020doubly}'s DR DiD estimand for the ATT can be biased when Assumption \ref{ass::stationarity} is violated. In contrast, our proposed estimand $\tau_{dr}$ is fully robust against compositional changes.

Proposition \ref{prop::bias} also highlights that not all violations of Assumption \ref{ass::stationarity} result in biases in ATT when using \citet{Sant2020doubly}'s estimand. Although intuitive and simple, this insight seems to be new in the literature. Based on this observation, one can determine if violations of Assumption \ref{ass::stationarity} lead to empirically relevant biases in the ATT by comparing nonparametric estimates based on $\tau_{sz}$ with those based on our proposed estimand $\tau_{dr}$. This would detect only the ``relevant'' violations of Assumption \ref{ass::stationarity} that affect the target parameter of interest. That is, it would concentrate power in the directions that one cares about in this context. We discuss this testing procedure in greater detail in Section \ref{sec::test}.

At this point, one may also wonder what the price one pays for such robustness in terms of semiparametric efficiency. Specifically, how much efficiency one loses by using  $\tau_{dr}$ when Assumption \ref{ass::stationarity} holds but is not fully exploited. The next proposition compares the semiparametric efficiency bound derived in Theorem \ref{thm::seb} with the one derived by \citet{Sant2020doubly}. 

\begin{proposition}(Efficiency loss under stationarity) \label{prop::eff.loss}
Suppose that Assumptions \ref{ass::sampling}, \ref{ass::ident}, and \ref{ass::stationarity} hold. Then
	\begin{equation}
 \rho_{sz}  \equiv	\sE[\eta_{\text{eff}}(W)^{2}] - \sE[\eta_{sz}(W)^{2}] = \dfrac{1-\sE[T]}{\sE[D] \sE[T] } \var{\tau(X) | D = 1},
	\end{equation}
\end{proposition}

It is evident from Proposition \ref{prop::eff.loss} that our proposed estimator is asymptotically less efficient than the one proposed by \citet{Sant2020doubly} when there are no compositional changes over time. The efficiency loss is greater if any of the following three quantities is larger:  1) the population ratio of the pre-treatment period vs. the post-treatment period, 2) the population proportion of the comparison group vs. the treated group,  and 3) the expected variability of treatment effect heterogeneity among the treated. In the extreme case where the treatment effect on the treated is homogeneous, our ATT estimator would achieve the same efficiency level as the one that imposes stationarity \textit{a priori}. However, we imagine this case is not very realistic.

Propositions \ref{prop::bias} and \ref{prop::eff.loss} characterize a bias-variance trade-off. Although our proposed estimand for the ATT is robust against Assumption \ref{ass::stationarity}, there is an asymptotic efficiency loss of not exploiting Assumption \ref{ass::stationarity} when it does hold.  We revisit this trade-off in Section \ref{sec::test}.

\section{Estimation and inference} \label{sec::estimation}
The results from Section \ref{sec::id_and_efficiency} suggest one can estimate the ATT by building on the EIF derived in Theorem \ref{thm::seb}, as emphasized by \eqref{eq.dr}. The results from Propositions \ref{prop::bias} and \ref{prop::eff.loss} also suggest a testing procedure to assess whether compositional changes translate to biased ATT estimates. However, all the discussions so far have involved estimands that depend on unknown nuisance functions, and we have not yet discussed how one can estimate these to form feasible two-step estimators. This section discusses how to proceed when adopting a fully nonparametric approach, avoiding additional functional form assumptions.

We first present a generic result that emphasizes that estimators based on \eqref{eq.dr} possess a \emph{rate} DR property, regardless of how the nuisance functions are estimated nonparametrically. Since we employ a nonparametric estimation procedure, model misspecifications are not a primary concern, at least asymptotically. This suggests that the traditional notion of DR estimators that leverage potentially misspecified parametric working models, is unsuitable for our procedure. However, we recognize that different nonparametric estimators may exhibit varying convergence rates for the nuisance parameters. Instead of requiring that both nonparametric models for the nuisance parameters converge to their true values sufficiently fast, i.e., the differences are $o_p(n^{-1/4})$, our rate DR property relies on weaker conditions that allow for a trade-off between the convergence rates of the two nuisance estimators. Heuristically, the rate DR property implies that nonparametric DiD estimators for the ATT based on the EIF are $\sqrt{n}$-consistent and asymptotically normal even if one of the outcome regression or generalized propensity score functions is very complex provided that the other is sufficiently simple. For more discussions on rate DR, see, e.g., \citet{Kennedy2016, Kennedy_2023_DR_review}, \citet{Smucler2019}, \citet{Rotnitzky2021}, \citet{Jordan_etal_2022_Neurips}, and \citet{Bonvini_etal_2024}.

Although interesting and useful, this generic rate DR is agnostic about the choice of the nonparametric estimator for the nuisance functions, and, therefore, does not help us with practical inference procedures. Towards that end, we discuss how one can concretely estimate the generalized propensity score (PS) and outcome regression (OR) nuisance functions using local polynomials, even in the presence of discrete covariates, in Subsection \ref{sec:est.lp}. We then establish the large sample properties of our DR DiD two-step estimator for the ATT based on local polynomials. We provide a data-driven bandwidth selection method in Subsection \ref{sec:bw}. We defer the construction of the Hausman-type test for compositional changes to Section \ref{sec::test}.

\subsection{Rate doubly robust\label{sec:rateDR}}
 
Let $ \phat$ be a generic estimator of $p$, and $\mhat_{d,t} $ a generic estimator of $m_{d,t}$ for $\dtnpt$. Given these first-step estimators, our proposed (generic) two-step estimator for the ATT based on  \eqref{eq.dr} is given by
\begin{align}
\tauhat_{dr} &= \sE_n\left[ \what_{1, 1}(D, T)\tauhat(Y, X) +  \sum_{\dtnpt}   (-1)^{(d + t)}  \what_{d,t}(D, T, X)  (Y - \mhat_{d,t}(X))\right],\label{eq.tau.dr.est}
\end{align}
where $\tauhat(Y, X) = Y - (\mhat_{1,0}(X) + (\mhat_{0,1}(X) - \mhat_{0,0}(X)) )$, and, for $\dtnpt$,
\begin{align}
	\what_{1,1}(D, T)  &=  \dfrac{DT }{\sE_{n}[DT]} ,  \label{eq.def.what1}  \\
	\what_{d, t}(D, T, X) & = \left. \dfrac{I_{d,t} \cdot \phat(1,1,X) }{\phat(d, t, X)}\middle / \sE_n\bracks{\dfrac{I_{d,t} \cdot \phat(1,1,X) }{\phat(d, t, X)}}\right. \label{eq.def.what2}.
\end{align}

Let  $ \norm{f}_{L_2} \equiv \parens{ \int f^2 d \mu  }^{1/2} $ and $\norm{f}_{\infty} \equiv \sup_{x\in\Xcal} \normabs{  f(x)  }$ denote the $L_2$- and sup-norm of a function $ f $, respectively, and let $ \sG_{n}(\cdot) $ denote the empirical process $ \sqrt{n}\parens{ \sE_{n} - \sE} (\cdot) $. We impose the following assumptions on the quality of nuisance function estimators.  

\begin{asm}[Estimation of nuisance parameters] \label{ass::genr.conv.rate}
\phantom{a}
\begin{enumerate}
  \item The estimators $ \phat $ and $ \mhat $ are uniformly consistent in the sense that
  \begin{align*}
   \norm{\phat(\cdot,\cdot, \cdot)  -  p(\cdot,\cdot,\cdot) }_{\infty} = \op(1),  \quad \max_{\dtnpt} \norm{\widehat{m}_{d,t}(\cdot)  -  m_{d,t}(\cdot) }_{\infty} = \op(1).
 \end{align*}

\item For a positive sequence $(\varepsilon_n^{\ast})_{n \geq 1}$ with $\varepsilon_n^{\ast}\to 0$, the nuisance estimators $ (\phat, \{\mhat_{d,t}\}_{\dtnpt})$  lie in the function class $\Fcal \equiv \Fcal^p \times (\mathcal{F}^{m})^3 $ with probability at least $ 1- \varepsilon^{\ast}_n$. The sets $ \Fcal^{p} $ and $  \Fcal^{m}$  contain the true nuisance functions and satisfy the following constraints:
\begin{enumerate}[label=$(\roman*)$]
    \item  For all $ \dtall $ and $ \ptilde \in \Fcal^{p} $, $ 0 < \inf_{x \in \Xcal}  \ptilde(d,t, x)  <  \sup_{x \in \Xcal}  \ptilde(d,t, x) <1  $.
    \item For all $ \dtnpt $, $  \sup_{\mtilde_{d,t} \in \Fcal^{m}}\norms{\mtilde_{d,t}} < \infty$.
    \item Both $ \Fcal^{p} $ and $ \Fcal^{m}$ have finite uniform entropy integrals (see Appendix \ref{sec:notation} for the definition).
\end{enumerate}

    \item $\var{Y|D = 1, T = 1} < \infty $ and $ \sup_{x\in\Xcal}\var{Y|D = d, T = t, X = x}< \infty$ , for all $ \dtnpt.$

\end{enumerate}
\end{asm}

Assumption \ref{ass::genr.conv.rate}.2 places a mild restriction on the complexity of the function classes in which the nuisance estimators and their probability limits reside. Specifically, the requirement of a finite uniform entropy integral ensures that these function classes are not excessively rich while still allowing for a broad range of data-adaptive estimators without requiring parametric models. Many commonly used function classes satisfy this condition, including standard parametric families, smooth functions with uniformly bounded derivatives, and Lipschitz transformations of such classes; see, e.g., \citet{VanderVaart1996}, \citet{VanderVaart1998}, and \citet{Kosorok2008} for additional examples. It is possible to relax this condition, at the cost of invoking higher-level conditions; see Remark \ref{rem:equicontinuity} for a discussion

Let $ (r_n)_{n\geq 1}$ and $ (s_n)_{n\geq 1}$ be positive sequences converging to zero such that
 \begin{align}
  &\max_{\dtall} \norm{\phat(d,t, \cdot)  -  p(d,t,\cdot) }_{e} = O_{p}(r_{n}),  \label{eq.phat.conv.e.rate} \\
  &\max_{\dtnpt} \norm{\widehat{m}_{d,t}(\cdot)  -  m_{d,t}(\cdot) }_{e} = O_{p}(s_{n}), \label{eq.mhat.conv.e.rate}
 \end{align}
 where $ e  = L_2$ or $ \infty$. 

\begin{lem}[Doubly robust error rate with generic first step estimators] \label{lem::genr.conv.rate}
Under Assumptions \ref{ass::sampling}, \ref{ass::ident}, and \ref{ass::genr.conv.rate}, 
	\begin{equation} \label{eq.lem.1}
		\widehat{\tau}_{dr}  - \tau =  \dfrac{1}{n}\sum_{i = 1}^{n} \eta_{\text{eff}}(W_{i}) + \Op\parens{ r_{n} s_{n} } + \opnsq,
	\end{equation}
where the convergence rates $r_{n}$ and $s_{n} $ from \eqref{eq.phat.conv.e.rate}–\eqref{eq.mhat.conv.e.rate} hold for either $ e = L_2 $ or $ e = \infty $.
\end{lem}

The lemma demonstrates that our estimator is doubly robust in terms of its convergence rate. The remaining term is the product of the error rates of the first-stage estimators.  Due to the product structure, each estimator typically needs only to converge to its true value at a rate of $ o(n^{-1/4})$ for the ATT estimator to converge at the parametric rate. This property also allows for a trade-off between precision in the two nuisance estimators.

\begin{rem}\label{rem:equicontinuity}
A similar result can be established without invoking the uniform entropy conditions in Assumption \ref{ass::genr.conv.rate}.2, by working directly with stochastic equicontinuity conditions. This alternative approach, formalized in Lemma \ref{lem:conv.rate.equicont} in Appendix \ref{sec:prof.ident}, is useful when verifying uniform entropy conditions is technically challenging or when the complexity of the nuisance function classes is difficult to characterize explicitly. $\blacksquare$
\end{rem}



\subsection{Local polynomial estimation of nuisance functions} \label{sec:est.lp}
 
 We first introduce the estimator for the PS functions. Conditional probability functions are naturally bounded within the unit interval. However, these bounds may not be respected when using linear probability models. As a nonparametric generalization of parametric multinomial logit regression, local multinomial logit regression enforces such bounds by design. Through extensive Monte Carlo simulations, \citet{frolich2006non} demonstrates that the local multinomial logit estimator consistently outperforms local least squares, Klein–Spady, and Nadaraya–Watson estimators.  Hence, we prefer this estimator over other nonparametric methods. 
  
Without loss of generality, the generalized PS can be represented by a multinomial logistic transformation applied to a set of unknown functions $ \{  g_{d,t}(\cdot) \}_{\dtnpt} $ as follows
 \[   p(d,t,x)  =   \dfrac{\exp(g_{d,t}(x))}{ 1 + \sum_{\dtnptp} \exp(g_{d',t'}(x))   }, \]
 for $ \dtnpt  $, and $ p(1,1,x)  = \parens{1 + \sum_{\dtnptp} \exp(g_{d',t'}(x))  }^{-1}$. The representation is well-defined as long as the overlapping condition in Assumption \ref{ass::ident} (iii) holds. Instead of imposing specific functional forms on $ \{  g_{d,t}(\cdot) \}_{\dtnpt} $, the local multinomial logit estimator approximates these unknown functions locally using polynomials, which we will describe in detail below. 

In accordance with the conventions of local polynomial estimation, we adopt the following notations as shorthand for common vector operators,
\begin{align*}
 &\k = (k_{1}, ..., k_{v}),  \quad |\k| = \sum_{\ell = 1}^{v} k_{\ell}, \quad  \k ! = \prod_{\ell = 1}^{v}k_{\ell}!, \quad x^{\k } =\prod_{\ell = 1}^{v} x_{\ell}^{k_{\ell}} ,\\
  & f^{(\k)}(x) = \dfrac{\partial^{\k} f(x)}{ \partial x_{1}^{k_{1}} \cdot \partial x_{2}^{k_{2}}\cdots\partial x_{v}^{k_{v}} }, \quad  \sum_{0 \leq \normabs{\k} \leq p}   = \sum_{ \ell = 0}^{p} \underset{k_{1} +... + k_{v} = \ell}{\sum_{k_{1} = 0}^{\ell} ... \sum_{k_{v} = 0}^{\ell}  }.
\end{align*}

Furthermore, we define $ n_{k} = \binom {k + \ell - 1} {\ell - 1} $ as the number of distinct $  \ell $-tuples $ \k $ with $ \kabs = k $. We arrange these $ n_{k} $ $  \ell $-tuples in a lexicographically-ordered sequence, prioritizing the last position, and denote the mapping from the rank in the ordered sequence to the corresponding $  \ell $-tuple as $ \pi_{k}(\cdot)  $.

Since our method accommodates both discrete and continuous covariates, we must distinguish between these types of variables. We assume that $ x = (\xc, \xd) $, where $ \xc $ is a $ \xcdim $-vector of continuous covariates, and $ \xd $ is the subvector of discrete variables. We also distinguish between ordered and unordered discrete variables. That is, $ x_{d} = (\xu, \xo) $, where $ \xu $ is a $ \xudim $-vector of unordered covariates and $ \xo $ is a $ \xodim $-vector of ordered covariates.  

 Now, for a generic function, $ g: \Xcal \to \R $, and a point, $ x^{\ast} \in \Xcal $,  $ g(\cdot) $ can be approximated in a neighborhood of $ x^{\ast} $ by a $p$-th order Taylor series with respect to the continuous variables, as  
\begin{equation*}
	g(x)	 \approx   \sum_{ 0 \leq \kabs \leq p} \dfrac{1}{\kec} \deriv{g}{\k}{}(x^{\ast}) (x_{c} - x^{\ast}_{c})^{\k}  =  \bZ(x^{\ast}_{c}) ' \gamma_{g}(x^{\ast}),
\end{equation*}
where $ \bZ(x_{c}) = (\bZ^{(0)\transpose}(x_{c}) , ..., \bZ^{(p)\transpose}(x_{c}) )^{\transpose}  $ is a $ N_{p} \times 1 $ vector that contains the sorted $  (X_{c} - x_{c})^{\k}  $, with $N_{p} \equiv \sum_{k = 0}^{p}  n_{k}$. The $ l $-th entry of $ \bZ^{(k)}(x_{c}) $, denoted as $  \bZ^{(k, l)}(x_{c})$, is equal to $ (X_{c} - x_{c})^{\pi_{k}(l)} $. The vector $ \gamma_{g}(x)  = (\gamma_{g}^{(0)'}(x), \dots, \gamma_{g}^{(p)'}(x))'$ is defined as the vector of lexicographically-ordered $ \deriv{g}{\k}{}(x) /\kec $.

The local approximation is achieved through kernel smoothing.  For continuous variables, we let the kernel function be denoted by  $ K^{j}(\u) $, $ j = ps, or $.  It is a nonnegative function supported on $ [-1, 1]^{  \xcdim} $. Suppose $ h > 0 $ is a generic bandwidth parameter. We denote the scaled kernel function by $ 	K_{h}(\u) = K\parens{\u/h}/h^{\xcdim}$. We use the kernel function proposed by \citet{li2007nonparametric} for discrete variables.  This kernel function is defined as 
\begin{equation}\label{eq.kernel.li}
	 L_{\lambda}(x_{d}, z_{d}) =  \prod_{ s = 1 }^{\xudim}  \lambda_{u}^{\idx{ x_{u, s} \neq z_{u, s} }} \cdot \prod_{ \ell = 1 }^{\xodim}  \lambda_{o}^{\normabs{ x_{o, \ell} - z_{o, \ell} }},
\end{equation}
where $ \lambda = (\lambda_{u}, \lambda_{o}) \in [0, 1]^{  2} $ is a generic smoothing parameter.
When $ \lambda = (0, 0) $, the estimator reduces to the frequency estimator.\footnote{Here, we adopt the convention that $0^0 = 1$, ensuring the estimator remains well-defined even when $x_d = z_d$.}

For the $j$-th observation of covariates, $ X_{j} $, our local polynomial (multinomial) logit estimator of $\gamma$, denoted by $ \gammahat $, satisfies
\begin{equation} \label{eq.loc.lik}
    \gammahat(X_{j})  \equiv ( \gammahat'_{1,0}(X_{j}),  \gammahat'_{0, 1}(X_{j}), \gammahat'_{0, 0}(X_{j}))' = \argmax_{ \gamma \in \R^{3N_{p}} }\dfrac{1}{n-1} \sum_{i \neq j }^{n} \ell(W_{i}, X_{j};   \gamma) \kerps(X_{i}; X_{j}, h, \lambda),
\end{equation}
where $ \kerps(X_{i}; X_{j}, h, \lambda)  =   K_{\hp}^{ps}\parens{X_{c, i} - X_{c, j}} L_{\lambda}(X_{d}, X_{d,j})   $ and the local likelihood function $ \ell(w, x ;   \gamma)  $ is defined as
\begin{align*}
	\ell(w, x ;   \gamma) = \sum_{ \dtnptp } \dt \bZ_{p}(x_{c})^{\transpose}  \gamma_{d,t}    - \log\parens{1 +\sum_{\dtnptp} \exp\parens{ \bZ_{p}(x_{c})^{\transpose}  \gamma_{d',t'} } }.
\end{align*}
Note that we have used a ``leave-one-out'' version of the local regression estimator for the construction of $ \gammahat $, i.e., $ \gamma(X_{j}) $ are estimated using every observation except the $ j $-th.  This technique, standard in the literature \citep{powell1996optimal, powell1989semiparametric, rothe2019properties}, serves to avoid a ``leave-in'' bias that is of first-order importance when estimating the ATT. 

 Let $e_{\ell, k}$ denote an $\ell$-dimensional vector in which the $k$-th element is set to one, while all remaining elements are zero. Then, for a given $ \gammahat $, the generalized PS can be approximated by\footnote{We abuse notation and denote the local polynomial estimators for the generalized propensity score as $\phat$ and for the outcome regression as $\mhat$, which are the same as the generic estimators introduced in Section \ref{sec:rateDR}.}
\begin{equation}\label{eq.phat}
	\phat(d, t, x) =  \dfrac{\exp(\ep'  \gammahat_{d,t}(x))}{1 + \sum_{ \dtnptp } \exp(\ep' \gammahat_{d',t'}(x))}, 
\end{equation} 
 for $ \dtnpt $, and  $ 	\phat(1, 1, x)  =  1 - \sum_{\dtnpt} \phat(d, t, x)$.

For OR models, we employ leave-one-out $q$-th order local polynomial least squares estimators. First, the local polynomial regression coefficients are estimated by solving the following equation:
\begin{equation}\label{eq.mhat}
	\betahat_{d,t}(X_{j}) =  \argmin_{ \beta \in \R^{N_{p}} } \dfrac{1}{n-1}\sum_{ i \neq j }^{n} \parens{ Y_{i} - \bZ_{q, i}(X_{c,j})^{\transpose}  \beta}^{2} \dti  \keror(X_{i}; X_{j}, b_{d,t}, \vartheta_{d,t} ),
\end{equation}
where $ \keror(X_{i}; X_{j}, b_{d,t}, \vartheta_{d,t} )  = K^{or}_{b_{d,t}}\parens{X_{c, i} - X_{c, j}}  L_{\vartheta_{d,t}}(X_{d}, X_{d,j})  $, and $\dti = \idx{D_i=d, T_i = t}$. Then, we estimate the OR functions by 
\begin{equation}\label{eq.mrhat}
	\mhat_{d, t}(X_{j}) =   \eq^{\transpose} \betahat_{d,t}(X_{j}),
\end{equation} for $ \dtnpt $. 

We analyze the asymptotic behaviors of these local polynomial estimators in Appendix \ref{sec:aux.est}. We provide results on the uniform convergence rate for the approximation error. In particular, we establish a uniform stochastic expansion for the local multinomial logit regression that is of independent interest. 

\begin{rem}
	The choice of polynomial order depends on considerations such as computational tractability and the trade-off between bias and variance properties. We adhere to the recommendation made by \citet{fan1995local} to employ odd-degree polynomial fits, as they simplify the analysis for the boundary bias when using symmetric kernel functions. We allow varying local polynomial orders for the PS and OR estimators and, in the case of the latter, for distinct treatment groups.  This flexibility is desirable as the propensity score and conditional mean functions might display varying degrees of smoothness. $\blacksquare$
\end{rem}

\subsection{Asymptotic normality}
With $\{\mhat_{d,t}\}_{\dtnpt}$ given in \eqref{eq.mrhat}, and $\phat$ defined in \eqref{eq.phat}, we can construct an estimator for $\tau_{dr}$ as shown in \eqref{eq.tau.dr.est}. In the following, we derive the large sample properties of the estimator $ \tauhat_{dr} $ by applying Lemma \ref{lem::genr.conv.rate}. 
To achieve this objective, we begin by presenting a set of regularity assumptions. Henceforth, we use $\neigb{x, \delta}$ to denote a ball centered at $x$ with radius $\delta$, and  $\lambda_{min}(A)$ to represent the smallest eigenvalue of a square matrix $A$.

\begin{asm}[Support, smoothness, integrability, kernel, and bandwidth] \label{ass::asy.lp}
\phantom{a}
\vspace{-.3cm}
	\begin{enumerate}
		\item (i)  $ \Xcal = \Xcal_{c} \otimes \Xcal_{d} $, where $ \Xcal_{c} $ is a compact subset of $ \mathbb{R}^{\xcdim} $ and $ \Xcal_{d} $ is finite; (ii)  For all $ x_{d} \in \Xcal_{d} $, $ \prob{X_{d} = x_{d}} >0   $, and the conditional probability density of $ X_{c} $, $ f_{X_{c}|X_{d}}(\cdot| x_{d}) $, is continuously differentiable and bounded away from zero on $ \Xcal_{c} $; (iii) There are positive constants $\kappa_0 $ and $\kappa_1 $ in $ (0, 1] $ such that for any $ x \in \Xcal $ and all $ \epsilon \in (0, \kappa_0]  $, there exists a  $ x' \in \Xcal $ satisfying, $  x_d'= x_d $, and $\neigb{x', \kappa_1 \epsilon }  \subset  \neigb{x, \epsilon } \cap \Xcal.$

		\item For all $ x \in \Xcal $, (i) $ p(d,t,x) $ is  $ (p+1) $-times continuously differentiable in $ x_{c} $, with uniformly bounded derivatives, for $ \dtall $; (ii)  $ m_{d, t}(x) $ is  $ (q  +1) $-times continuously differentiable in $ x_{c} $, with uniformly bounded derivatives, for $ \dtnpt $.
	    \item $ \sE[|Y|^{\zeta} | X, D, T] < \infty $ a.s. for some constant $ \zeta  > 2. $ 
		\item For $ j = ps, or $, (i) $ K^{j} : [-1, 1]^{  \xcdim} \to \mathbb{R}_{+}$; (ii) $ K^{j}(\cdot) $ satisfies the Lipschitz condition, i.e. $ \normabs{ K^j(\u) -  K^j(\u')} \leq L \norm{\u - \u' } $  for some $ L >0 $ and any $ \u, \u' \in \R^d $.  
		\item  (i) $ \hp = \oh(1) $; 
		(ii)  $ \log n/\parens{n h^{v_c + 2p}} = o(1)  $ and $\lambda/h^p  = o(1)$; 
		(iii)  $  \bhp = \oh\parens{n^{-1/4}}  $ and  $  \log n / \parens{ n \php^{\xcdim}} = \oh\parens{n^{-1/2}}  $.
		 For  $ \dtnpt $, (iv)  $ b_{d,t} = \oh(1) $;  
		 (v)  $   \log n / \parens{n^{1-2/\zeta} b_{d,t}^{\xcdim}} =\oh(1) $; 
		 (vi)  $  \bhor = \oh\parens{n^{-1/4}}  $ and  $  \log n / \parens{ n \phor^{\xcdim}} = \oh\parens{n^{-1/2}}  $;   
		 (vii) $\lambda, \vartheta_{d,t} = \oh(n^{-1/4}) $.
   \item  With $\Qx{j}$ defined in \eqref{eq.q.def}, $ \inf_{x_{c}\in \Xcal_{c}}\lambda_{min}\parens{\Qx{j}} > 0 $, for $j = p, q$.
	\end{enumerate}
\end{asm}

A few remarks on the assumptions are in order. Assumption \ref{ass::asy.lp}.1 indicates that our local polynomial estimator can handle discrete, categorical data. The final part of the condition, proposed by \citet{Fan2016}, requires that the boundary of $\Xcal$ is sufficiently dense for the first-stage estimators to exhibit good bias and variance properties near the boundary. Importantly, this has to hold for all covariates.\footnote{If this denseness condition does not hold in practice---for example, if the data is very sparse near the boundary of covariates---it may be necessary to restrict the analysis to interior points and impose trimming when estimating treatment effects.}
Assumption \ref{ass::asy.lp}.2 describes the standard smoothness condition for the nuisance functions.  Assumption \ref{ass::asy.lp}.3 is a regularity condition that controls the conditional moments of $ Y $.  Assumption \ref{ass::asy.lp}.4 collects the regularity conditions on the kernel functions. We note that different kernels can be used for the propensity score and conditional mean models. In practice, the kernel $K(\cdot) $ typically takes a product form, that is, $ K(\u) = \prod_{i = 1}^{\xcdim} \mathcal{K}(u_{i}) $, where $ \mathcal{K}(\cdot) $ can be chosen from several options, such as triangular, biweight, triweight, or Epanechnikov kernels. However, the Gaussian kernel is ruled out due to the restriction on compact support. Assumption \ref{ass::asy.lp}.5 compiles the rate condition on the bandwidths. Assumptions \ref{ass::asy.lp}.5 (ii) and (v) are imposed to ensure linear expansions of the local polynomial estimators hold uniformly over $ \Xcal $.  When $ Y $ has finite moments of any order, such as when it has bounded support, Assumption \ref{ass::asy.lp}.5 (v) is implied by Assumption \ref{ass::asy.lp}.5 (vi).  Assumptions \ref{ass::asy.lp}.5 (iii), (vi), and (vii) specify rate conditions on the bias and stochastic part of the first-step estimation error. 

It is important to note that our estimator builds on the efficient influence function and, therefore, inherits a rate DR property. Without such a DR property, it would typically require more stringent rate conditions on the bias part, which can only be satisfied with higher-order kernel functions. Heuristically, this follows because one needs to ensure that the nonparametric estimator converges fast enough. See, for example, \citet{Newey1994} and \citet{Lee2018a} for detailed discussion. The rate DR property relax this condition.

\begin{rem} \label{rem:conv.rate}
\citet{rothe2019properties} provides a result that can be applied to weaken the rate conditions on the nuisance functions. They present higher-order expansions of semiparametric two-step DR estimators, demonstrating that if the first-step error's bias and the stochastic components are of order $ \op(n^{-1/6})  $, and their product is of order $ \op(n^{-1/2}) $, the resulting DR estimator achieves root-$ n $ consistency.  To maintain focus, we will not delve into an in-depth discussion on this topic. $\blacksquare$
\end{rem}

\begin{thm} (Asymptotic normality doubly robust estimator) \label{thm::asy.normal}
	Under Assumptions \ref{ass::sampling}, \ref{ass::ident}, and \ref{ass::asy.lp}, we have
	\begin{equation}
	\sqrt{n}(\widehat{\tau}_{dr}  - \tau) =   \dfrac{1}{\sqrt{n}}\sum_{i = 1}^{n} \eta_{\text{eff}}(W_{i})  + \op(1) \convd \normnot{0}{\Omega_{dr}},
	\end{equation}
where $ \Omega_{dr}  =  \sE[\eta_{\text{eff}}(W)^{2}] $.
\end{thm}

Theorem \ref{thm::asy.normal} states that $ \tauhat_{dr} $ is root-$ n $ consistent, and asymptotically normal. It also shows that the estimation error of the nuisance functions does not affect the asymptotic distribution of $ \tauhat_{dr} $. Furthermore, the asymptotic variance of $ \tauhat_{dr} $ is equal to the semiparametric efficiency bound.

The theorem can be applied to calculate confidence intervals for the ATT. To achieve this, we need an estimator of the asymptotic variance, $ \Omega_{dr} $. One approach to constructing such an estimator is by using empirical analogs of the influence function or through bootstrapping. Here, we focus on the first method, while a weighted bootstrap procedure that accommodates clustered inference is provided in Appendix \ref{sec:boot}. Let
\begin{equation} \label{eq.etahat}
	\etahat_{\text{eff}}(W)  = \sum_{\dtnpt}(-1)^{d+t} \what_{d,t}(D, T, X) (Y - \mhat_{d,t}(X)) + \what_{1,1}(D, T)  (\tauhat(Y, X) - \tauhat_{dr}), 
\end{equation}
and $ \Omegahat_{dr}  =  \sE_{n}[\etahat_{\text{eff}}(W )^{2}]  $.  Under mild regularity conditions, the consistency of  $ \Omegahat_{dr} $ can be established, with its proof included in that of Theorem \ref{thm::test.hausman} presented in the following section.

\subsection{Bandwidth selection} \label{sec:bw}
This subsection addresses the practical selection of bandwidth for the first-step local polynomial estimators.  It is well-documented that smoothing parameters have a significant impact on balancing the trade-off between bias and variance. Although robustness checks employing multiple bandwidths can be useful, a reliable data-driven selection rule is often preferred.  In the following, we outline two cross-validation procedures for choosing these tuning parameters.

Define the following two criterion functions
\begin{align}
	C_{n}^{ls}(h, \lambda,  \{b_{d,t}, &\vartheta_{d,t} \}_{\dtnpt}   ) \nonumber \\
	          &  =    \sumi \cbracks{\sum_{ \dtall } (\dti - \phat(d,t, X_{i}))^{2} + \sum_{ \dtnpt }\dti (Y_{i} - \mhat_{d,t}(X_{i}))^{2} }, \label{eq.cv.ls}\\
	C_{n}^{ml}(h,  \lambda, \{b_{d,t}, &\vartheta_{d,t} \}_{\dtnpt}   ) \nonumber  \\
	          &  =    \sumi\cbracks{ -\sum_{ \dtall } \dti \log(  \phat(d,t, X_{i}))   + \sum_{ \dtnpt }\dti  (Y_{i} - \mhat_{d,t}(X_{i}))^{2} }. \label{eq.cv.ml}
\end{align}

The least-squares criterion, $ C_{n}^{ls}$,  is a standard choice in the kernel estimation literature. It is based on the sum of the least squares distances between the observed and leave-one-out fitted values for both PS and OR estimators, The second criterion, $ C_{n}^{ml} $, replaces the PS estimator's least squares sum with that of the observed likelihood.  This idea of using a likelihood-based criterion in local logistic estimation can be traced back to \citet{staniswalis1989kernel}. 

The leave-one-out cross-validated bandwidths, $ \parens{\hhat^{j}, \widehat{\lambda}^{j}, \{\bhat^{j}_{d,t}, \widehat{\vartheta}^{j}_{d,t}\}_{\dtnpt}} $, minimizes $ C_{n}^{j} $ for $ j = ls, ml $.  
In Appendix \ref{sec:mse}, we analyze the mean integrated squared error (MISE) properties of the first-step estimators and derive the convergence rates of the optimal bandwidths. For local linear estimation (i.e., 
$p=q=1$), Theorem \ref{thm:mse.bandwidth} shows that the optimal bandwidths ensure the rate conditions in Assumption \ref{ass::asy.lp}.5 are satisfied if the number of continuous variables is less than 4. Notably, this result imposes no restrictions on the number of discrete variables.

\begin{rem}
Leave-one-out cross-validation can be computationally demanding when combined with local multinomial logit estimation. This is partly because, unlike local least squares regression, local multinomial logit regression does not have a closed-form solution. As a result, evaluating the criterion function requires solving $n$ minimization problems, which can be time-consuming, especially for large datasets. To mitigate this computational burden, we propose using the rescaled cross-validation method introduced by \citet{li2021nonparametric}, as described in Appendix \ref{sec:rcv}. This method divides the data into training and validation sets and computes the multinomial logistic loss using only the validation data, significantly reducing the computation cost. $\blacksquare$
\end{rem}

\section{Testing for compositional changes} \label{sec::test}

Propositions \ref{prop::bias} and \ref{prop::eff.loss} reveal that our proposed estimator for the ATT is robust against compositional changes; however, it is less efficient than the DR DiD estimator proposed by \citet{Sant2020doubly} when the covariate-stationarity assumption is correctly imposed. This trade-off suggests a nonparametric \citet{hausman1978specification}-type test for the absence of compositional changes can be constructed by comparing our proposed estimator with that of \citet{Sant2020doubly}. Although \citet{Sant2020doubly} focuses on parametric first-step estimators for the nuisance parameters, we modestly extend their analysis by considering nonparametric first-step estimators in this section. 

Before detailing the test construction, we define the null and alternative hypotheses. Let $  \Pb_0 = \{ P \in \Pcal: \text{Assumptions \ref{ass::sampling}, \ref{ass::ident}, and \ref{ass::stationarity} hold, and }   \varp{\tau(X) \mid D = 1} >0  \}$ and $\Pb_1 = \{ P \in \Pcal: \text{Assumptions \ref{ass::sampling} and \ref{ass::ident} hold, and } \tau_{dr} \neq \tau_{sz} \}$, where $\tau_{dr} $ and  $\tau_{sz} $ are defined in \eqref{eq.dr} and \eqref{eq.sz}, respectively. Here, we aim to test
\begin{eqnarray*}
\hnull : P \in \Pb_0 
\text{~~~~~~~~against~~~~~~~~} \halt :  P \in \Pb_1.
\end{eqnarray*}%
We index the hypotheses with \( P \) to emphasize the point-wise nature of our test---namely, that it applies to a fixed data-generating distribution \( P \in \Pcal \). Under the null hypothesis, \citet{Sant2020doubly}'s DR DiD estimand coincides with our proposed estimand, so that \(\tau_{sz} = \tau_{dr}\). In this setting, both estimands identify the ATT, though their estimator (asymptotically) achieves the semiparametrically efficient bound. Under $\halt$, \citet{Sant2020doubly}'s DR DiD estimator is not consistent for the ATT, while ours remains consistent and is (asymptotically) semiparametric efficient. Finally, we note that for distributions satisfying Assumptions \ref{ass::sampling} and \ref{ass::ident}, the alternative hypothesis space constitutes only a subset of the complement of the null hypothesis space. This occurs because the two ATT estimators can still coincide, even when the stationarity assumption is violated. We concentrate on this subset of hypotheses to test deviations where the stationarity assumption influences the target parameter of interest.

\begin{rem} \label{rem:bias.bound}
To assess how the difference between $\tau_{dr}$ and $\tau_{sz}$ reflects 
violations of Assumption~\ref{ass::stationarity}, we use Proposition~\ref{prop::bias}:
\begin{align*}
|\tau_{sz} - \tau_{dr}| 
&= \Pr(T=0\mid D=1)
   \big|\mathbb{E}[\tau(X)\mid D=1,T=1]
        - \mathbb{E}[\tau(X)\mid D=1,T=0]\big| \\
&\le \Pr(T=0\mid D=1)
   \Big( 2 B_\tau \cdot TV(P_{1}, P_{0}) \Big) \\
&\le \Pr(T=0\mid D=1)
   \Big( 2 B_\tau \cdot \sqrt{\tfrac12 
        \min\{D_{\mathrm{KL}}(P_{1}\|P_{0}),
             D_{\mathrm{KL}}(P_{0}\|P_{1})\}} \Big),
\end{align*}
where $B_{\tau} =  \sup_{x \in \Xcal}\normabs{\tau(X)} $, $TV(P_1, P_0)$ and $D_{\mathrm{KL}}(P_1 || P_0)$ denote the total variation and Kullback-Leibler (K-L) divergence between $P_1 $, the distribution of $X|D=1,T=1$, and $P_0 $, the distribution of $X|D=1,T=0$, respectively.\footnote{For two probability measures $P$ and $Q$ defined on the same measurable space and dominated by a common $\sigma$-finite measure $\mu$, the total variation distance and the Kullback--Leibler divergence are defined as  $TV(P,Q) = \sup_{A} |P(A)-Q(A)|  = \tfrac12 \int |p-q|\,d\mu$ and $D_{\mathrm{KL}}(P\|Q) = \int p\log(p/q)\,d\mu$, where $p = dP/d\mu$ and $q = dQ/d\mu$.} The last line follows from Pinsker’s inequality, showing that the difference between these two estimands is bounded above by the extent of covariate shift as measured by standard distributional distances. A natural consequence of this simple observation is that tests for compositional changes that are constructed by comparing  $\tau_{sz}$ and $\tau_{dr}$ are likely to have limited power when $\big|\mathbb{E}[\tau(X)\mid D=1,T=1] - \mathbb{E}[\tau(X)\mid D=1,T=0]\big|$ or the relative sample size sampled in pre-treatment period among treated is small. One may argue that this is not necessarily an issue for learning about ATT in DiD setups with repeated cross-sections, as estimates of $\tau_{sz}$ and $\tau_{dr}$ are likely to be close despite the compositional changes assumption being violated. To some extent, our specification test rejects
Assumption \ref{ass::stationarity} when such violations ``impact'' ATT.

To illustrate this point, suppose $X|D=1,T=j \sim \mathcal{N}(\mu_j,\Sigma_j)$, $j=0,1$. We analyze whether $|\tau_{sz}-\tau_{dr}|=0$ in two stylized conditional average treatment effects functions:
\begin{enumerate}
  \item {Linear CATT:} 
  If $\tau(X)=\beta_0 + \beta_1^\top X$, 
  then $|\tau_{sz}-\tau_{dr}|=0$ if and only if $\beta_1^\top \mu_1=\beta_1^\top \mu_0$. 
  \item {Quadratic CATT:} 
  If $\tau(X)=\beta_0 + \beta_1^\top X + X^\top A X $, 
  then $|\tau_{sz}-\tau_{dr}|=0$ if and only if
  $\operatorname{tr}(A\Sigma_1)+\mu_1^\top A\mu_1+\beta_1^\top\mu_1
   = \operatorname{tr}(A\Sigma_0)+\mu_0^\top A\mu_0+\beta_1^\top\mu_0$.
\end{enumerate}

As the examples above highlight, even when $\mu_1\neq\mu_0$ and $\Sigma_1\neq\Sigma_0$ (so the KL divergence is positive), these ATTs estimands can coincide. For example, take $X$ to be univariate in the linear CATT with $\mu_0=\mu_1$, but $\Sigma_0 \ne \Sigma_1$: we have that $|\tau_{sz}-\tau_{dr}|=0$ even though KL divergene is positive. Alternatively, consider the quadratic CATT with $\mu_0=0$, $\mu_1 = 1$, $\beta_1 = 1$, $A=1$: as long as $\Sigma_0 = \Sigma_1 + 2$, it also follows that $|\tau_{sz}-\tau_{dr}|=0$, though the KL divergence is positive. Overall, one may argue that cases with $|\tau_{sz}-\tau_{dr}|=0$ in the presence of compositional changes are atypical, though we caveat that our assumptions do not rule that out. Our tests are not designed to capture such classes of alternatives. We expect practitioners to face a bias-variance trade-off in setups where $\big|\mathbb{E}[\tau(X)\mid D=1,T=1] - \mathbb{E}[\tau(X)\mid D=1,T=0]\big|$ is potentially small, so we recommend proceeding with caution. One could potentially construct alternative tests for Assumption \ref{ass::stationarity} based on the K-L divergence between $P_1$ and $P_0$, or attempt to adapt to potential model misspecifications in the sense of \citet{Armstrong_Kline_Sun_2024_adaptive}. We leave a detailed study of these topics for future research. $\blacksquare$
\end{rem}

To operationalize this testing procedure without invoking additional parametric assumptions, we require a nonparametric estimator for $\tau_{sz}$, which in turn necessitates nonparametric estimators for the PS $\tilde{p}(\cdot)$ and the OR functions $ m_{d,t}(\cdot)$, $\dtall$.  For the PS, we can use the local polynomial estimators from Section \ref{sec:est.lp} to construct an estimator for $\tilde{p}(\cdot)$ as
\[  \widehat{\tilde{p}}(X) = \phat(1,1, X) +   \phat(1, 0, X),  \]
where $ \phat(1, t, X) $ is given by \eqref{eq.phat}.  We can estimate the OR $ m_{d,t}(\cdot)$ as in \eqref{eq.mrhat}, noting that all four conditional mean functions must be estimated here (unlike the three in Section \ref{sec::estimation}).  Using these, we nonparametrically estimate $\tau_{sz}$ by
\begin{align} \label{eq.tau.sz.est}
\tauhat_{sz} &\equiv \sE_n\left[\dfrac{D}{\sE_n[D]}\tauhat(X) + \sum_{\dtall}   (-1)^{(d + t)}  \what^{sz}_{d,t}(D, T, X)  (Y - \mhat_{d,t}(X))\right].
\end{align}
 where  $ \tauhat(x) = (\mhat_{1,1}(x) - \mhat_{1,0}(x)) - (\mhat_{0,1}(x) - \mhat_{0,0}(x))$, and, for $t=0,1$,
\begin{eqnarray*}
\what_{1,t}^{sz}\left( D,T, X\right) &=&\frac{D\cdot \idx{T=t} }{\mathbb{E}_n\left[ D\cdot \idx{T=t} \right] }, \nonumber \\
\what_{0,t}^{sz}\left( D,T,X\right) &=&\left. \frac{\widehat{\tilde{p}}(X)\left( 1-D\right)
\cdot \idx{T=t} }{1-\widehat{\tilde{p}}(X)}\right/ \mathbb{E}_n\left[ \frac{\widehat{\tilde{p}}(X)\left( 1-D\right) \cdot \idx{T=t} }{1-\widehat{\tilde{p}}(X)}\right] .
\end{eqnarray*}

 Given this nonparametric estimator for $\tau_{sz}$ and our nonparametric estimator for $\tau_{dr}$ in \eqref{eq.tau.dr.est}, the test statistic is defined as
 \begin{equation}\label{eq.test.stat}
 	\Tcal_{n}  = n \Vhat^{-1}  \parens{ \tauhat_{dr} - \tauhat_{sz} }^{2},
 \end{equation}
 where $$ \Vhat \equiv \sE_{n}\bracks{\parens{ \etahat_{\text{eff}}(W) -   	\etahat_{sz}(W) }^{2}  }, $$  with $\etahat_{\text{eff}}(W)$ defined in \eqref{eq.etahat} and 
 \begin{equation}\label{eq.eta.sz.hat}
 	\etahat_{sz}(W) \equiv  \dfrac{D}{\sE_{n}[D]} ( \tauhat(X) -\tauhat_{sz} ) + \sum_{\dtall}   (-1)^{(d + t)}  \what^{sz}_{d,t}(D, T, X)  (Y - \mhat_{d,t}(X)).
 \end{equation}
$\Vhat$ estimates the variance of the difference between the two DiD estimators for the ATT. While an alternative variance estimator could be constructed using the variances of each DiD estimator, i.e., $\widetilde{V}_n = \Omegahat_{dr} - \Omegahat_{sz}$, with $ \Omegahat_{dr}  =  \sE_{n}[\etahat_{\text{eff}}(W )^{2}]  $ and $ \Omegahat_{sz}  =  \sE_{n}[\etahat_{sz}(W )^{2}]$, this approach may yield negative variance estimates in finite samples. Using $\Vhat$ avoids this drawback.

 In the following theorem, we characterize the asymptotic behavior of this statistic.  Let $ c^{\ast}_{1-\alpha} $ denote the $ (1-\alpha) $-th quantile of the chi-squared distribution with one degree of freedom (i.e. $ \chisq{1} $). 
 \begin{thm} \label{thm::test.hausman}
 	Suppose Assumption \ref{ass::asy.lp}, and in addition, Assumptions \ref{ass::asy.lp}.2(ii) and \ref{ass::asy.lp}.5(iv)-(vii) are fulfilled for $(d, t) = (1, 1) $, for $ P \in \mathcal{P}$.  Then:  \\
 	(a) Under the null hypothesis, $ \hnull $, we have  $ \Vhat \convp \rho_{sz} > 0  $, and 
 	\begin{equation}\label{eq.test.h0}
 		\lim_{n \to \infty } \prob{\Tcal_{n}  \geq  c^{\ast}_{1-\alpha}}  = \alpha;
 	\end{equation}
 	(b) Under the alternative hypothesis, $ \halt $, we have
 	\begin{equation}\label{eq.test.h1}
 		\lim_{n \to \infty } \prob{\Tcal_{n}  \geq  c^{\ast}_{1-\alpha}}  = 1.
 	\end{equation}
 \end{thm}

 The theorem states that the test controls size and is consistent. Although not discussed in detail here, it is easy to show that our test also has power against sequences of Pitman-type local alternatives that converge to the null at the parametric rate.

 \begin{rem}
  It is crucial to recognize that our test should be viewed as a  ``model validation'' instead of a ``model selection'' procedure. For researchers concerned about the validity of Assumption \ref{ass::stationarity},  it may be tempting to perform a two-stage test. In the first stage, a Hausman specification test is used to ``pretest'' for the presence of compositional changes, and then, in the second stage, the usual $t$-test is conducted based on either $ \tauhat_{dr} $ or $ \tauhat_{sz} $, depending on the outcome of the Hausman-test. However, as demonstrated by \citet{Guggenberger2010ET}, \citet{Guggenberger2010JoE}, and \citet{Roth2022}, such a model-selection procedure can lead to substantial size distortions when using standard inference methods. $\blacksquare$
 \end{rem}

\section{Monte Carlo simulation study}\label{sec::simulation}
In this section, we examine the finite sample properties of our proposed estimators and testing procedure. We conduct two main Monte Carlo experiments in this section.  In the first experiment, there are compositional changes over time, so Assumption \ref{ass::stationarity} is violated. In contrast, the second experiment adheres to this assumption, maintaining a joint distribution of covariates and treatment that is independent of treatment timing.  For each design, we compare our nonparametric DR DiD estimator $ \tauhat_{dr} $ defined in \eqref{eq.tau.dr.est}, which is robust against compositional changes and semiparametrically efficient, with the nonparametric extension of \citet{Sant2020doubly}'s estimator $\tauhat_{sz}$ defined in \eqref{eq.tau.sz.est}, which assumes no compositional change, and with the estimates of the regression coefficients, $ \tau_{fe} $, associated with two-way fixed effect (TWFE) regression specifications of the type 
$$Y = \alpha_{1} + \alpha_{2}T  + \alpha_{3} D  + \tau_{fe} (T\cdot D) + \theta ' X + \epsilon.$$ We consider two TWFE specifications: 1) a linear specification, where all the covariates $X$ enter linearly, and 2) a saturated specification, where, in addition to the linear terms, quadratic terms of the continuous covariates and all the interactive terms of the covariates are also included.  We include the TWFE specifications in our comparison set as they are prominent in empirical work.

We employ local linear ($ p, q = 1 $) kernel estimators for both the PS and OR functions.  As described in Section \ref{sec:est.lp}, the PS is estimated using the local likelihood method with the (multinomial) logistic link function, whereas the OR is estimated using the local least squares estimator. We utilize the second-order Epanechnikov kernel for the continuous covariates, and the kernel given in \eqref{eq.kernel.li} for discrete variables. Bandwidth selection methods are explored in detail later in this section.

Our main experiments involve a sample size of $ n  = 1000$, with each design undergoing $ 5,000 $ Monte Carlo replications. We evaluate the DiD estimators for the ATT using various metrics: average bias, median bias, root mean square error (RMSE), empirical 95\% coverage probability, the average length of a 95\% confidence interval, and the average of the plug-in estimator for the asymptotic variance. Confidence intervals are calculated using a normal approximation, with asymptotic variances estimated by their sample analogues. We also compute the semiparametric efficiency bound for each design to gauge the potential loss of efficiency/accuracy associated with using inefficient DiD estimators for the ATT. We perform a Hausman-type test as described in Section \ref{sec::test} under each design and report the empirical rejection rates. 

In addition to these two main experiments, we further investigate the power properties of our Hausman-type test by examining its performance under a sequence of local alternatives. Finally, we evaluate the performance of our estimator under different bandwidth selection methods, comparing leave-one-out cross-validation (LOOCV), rescaled cross-validation (RCV), and a plug-in estimator.

\subsection{Simulation 1: non-stationary covariate distribution} \label{sec:sim.1}
We first consider a scenario in which the stationarity condition is not satisfied.  The DGP is described in Appendix \ref{sec:sim.dgp}. Under this design, the covariate distribution does not exhibit time variation. However, the PS function is different in the two cross-sections.  The mean absolute difference between $ p^{s1}(1, 1, X) $ and $ p^{s1}(1, 0, X) $, as well as between $ p^{s1}(0, 1, X) $ and $ p^{s1}(0, 0, X) $, are both approximately $ 0.125 $, with the maximum difference reaching up to $ 0.63 $.\footnote{See Appendix \ref{sec:sim.dgp} for a definition of the $p^{s1}$ functions.}  Consequently, we expect all estimators to produce biased results, except for $ \tauhat_{dr} $. In addition, the stationarity test is likely to reject the null hypothesis with high probability. The results in Table \ref{tab:sim.tab.1} support these claims.

 \begin{table}[!ht]
 	\caption{Monte Carlo results under compositional changes. Sample size: $n = 1,000$.}%
 	\label{tab:sim.tab.1}%
 	\centering
 	\begin{adjustbox}{ width=0.8\linewidth, max totalheight=0.7\textheight, keepaspectratio}
 		\begin{threeparttable}
 \begin{tabular}{cccccccc}   
 	 \toprule   \multicolumn{8}{c}{True value of ATT: 4.31. Semiparametric Efficiency Bound: 1753.6} \\    \midrule          & \multicolumn{7}{c}{Two-way Fixed Effect Estimators} \\
   \noalign{\vskip 1mm} \cmidrule{2-8}          & Spec. & Avg. Bias & Med. Bias & RMSE  & Asy. Var. & Cover. & CIL \\
   \cmidrule{2-8}    $\tauhat_{fe}$ & Linear &-10.437 & -10.445 & 10.933 & 10425.033 & 0.121 & 12.633 \\ 
   $\tauhat_{fe}$ & Saturated &  -11.176 & -11.206 & 11.579 & 8797.289 & 0.045 & 11.612 \\  
   \cmidrule{2-8}  \noalign{\vskip 1mm}   & \multicolumn{7}{c}{Nonparametric Doubly Robust DiD Estimators for the ATT} \\
   \cmidrule{2-8}          & CV Crit. & Avg. Bias & Med. Bias & RMSE  & Asy. Var. & Cover. & CIL \\
   \cmidrule{2-8}    $\tauhat_{dr}$ & ML    &	 -0.009 & -0.010 & 1.374 & 1838.495 & 0.949 & 5.304 \\    
   $\tauhat_{dr}$ & LS    & -0.013 & -0.010 & 1.379 & 1848.848 & 0.949 & 5.314 \\    
   $\tauhat_{sz}$ & ML    & 4.427 & 4.436 & 4.543 & 983.436 & 0.009 & 3.884 \\    
   $\tauhat_{sz}$ & LS    & 4.427 & 4.435 & 4.543 & 983.746 & 0.009 & 3.884   \\   
    \cmidrule{2-8}  \noalign{\vskip 1mm}  \multicolumn{8}{c}{Hausman-type test} \\    
   \midrule    & CV Crit. & \multicolumn{1}{l}{Avg. Test Stats.} & \multicolumn{1}{l}{Emp. Pow. (0.10)} & \multicolumn{1}{l}{Emp. Pow. (0.05)} & \multicolumn{1}{l}{Emp. Pow. (0.01)} & & \\ 
   \midrule   & ML  &  21.250 & 0.998 & 0.996 & 0.978   & & \\ 
   &  LS    & 21.199 & 0.998 & 0.995 & 0.976  & & \\    
   
   \bottomrule    
 \end{tabular}
 			\begin{tablenotes}[para,flushleft]
 				\footnotesize{
 					Note: Simulations based on 5,000 Monte Carlo experiments. $ \tauhat_{fe} $ the TWFE regression estimator, $ \tauhat_{dr} $ is our proposed nonparametric DR DiD estimator \eqref{eq.tau.dr.est}, and $ \tauhat_{sz} $ is the nonparametric DR DiD estimator \eqref{eq.tau.sz.est} based on \citet{Sant2020doubly}. For TWFE regression, we use a linear specification, ``Linear'', and a saturated specification, ``Saturated''. For DR DiD estimators, the PS and the OR models are estimated using a local linear logistic and a local linear least squares regression, respectively.  Bandwidth for the PS function is selected with the log-likelihood criterion, ``ML'', and the least squares criterion, ``LS'', respectively.  Lastly, ``Spec.'', ``CV Crit.'', ``Avg. Bias'', ``Med. Bias'', ``RMSE'', ``Asy. Var.'', ``Cover.'', and ``CIL'', stand for the specification, cross-validation criterion,  average simulated bias, median simulated bias, simulated root-mean-squared errors, average of the plug-in estimator for the asymptotic variance, 95\% coverage probability, and 95\% confidence interval length, respectively.  The Hausman-type test statistic is calculated based on \eqref{eq.test.stat}.  Columns ``Avg. Test Stats.'', and ``Emp. Pow. $ (\alpha) $'' stand for the average test statistic, and empirical power of the test with a nominal size $ \alpha $, respectively. See the main text for further details.}
 			\end{tablenotes}
 		\end{threeparttable}
 \end{adjustbox}\end{table}

 First, results in Table \ref{tab:sim.tab.1} suggest that both $ \tauhat_{fe} $ and $ \tauhat_{sz} $ are severely biased under this DGP, while $ \tauhat_{dr} $ exhibits negligible bias on average.  Moreover, among the three sets of estimators considered, only our proposed estimator attains the correct coverage rate. This result is robust to the bandwidth selection method. Notably, the performance of the TWFE does not improve with a fully-saturated specification, indicating that incorporating nonlinear terms into a TWFE regression does not generally help in identifying heterogeneous treatment effects. In terms of efficiency, it is worth noting that the asymptotic variance of $\tauhat_{dr}$ is close to the semiparametric efficiency bound, which corroborates the findings of Theorem \ref{thm::asy.normal}. Regarding the testing performance, our Hausman-type test can effectively distinguish between the two nonparametric DiD estimators with a high degree of certainty, which is in line with our theoretical finding.

\subsection{Simulation 2: stationary covariate distribution}
 We now slightly adjust the first design by taking the average of propensity scores over time while keeping all other aspects of the DGP constant.  Specifically, we define
  \begin{align*}
 	p^{s2}(d, t, X) & =\mathbb{P}^{s1}\parens{T = t}\cdot (p^{s1}(d, 1, X) +  p^{s1}(d, 0, X)),
 \end{align*}
 where $\mathbb{P}^{s1}\parens{T = t} = \sE[p^{s1}(1, t, X)  + p^{s1}(0, t, X)]  $. The treatment groups are then assigned based on the realization of a standard uniform random variable on the unit interval partitioned by $ \{p^{s2}(d,t,X)\}_{\dtall} $.   Furthermore, the potential outcomes are determined by \eqref{dgp1.or.1}--\eqref{dgp1.or.3}. Unlike the first DGP,  both the covariate distribution and the  propensity score function are stationary in this case. As a result, we anticipate that both $ \tauhat_{dr} $ and $ \tauhat_{sz} $ will be consistent for the true ATT.  Furthermore, the empirical rejection rate of the Hausman-type test  is expected to  converge to the nominal sizes.  The Monte Carlo results under this DGP are summarized in Table \ref{tab:sim.tab.2}. 

 \begin{table}[!ht]
	\caption{Monte Carlo results under no compositional changes. Sample size: $n = 1,000$.}%
	\label{tab:sim.tab.2}%
	\centering
	\begin{adjustbox}{ width=0.8\linewidth, max totalheight=0.7\textheight, keepaspectratio}
		\begin{threeparttable}
		 \begin{tabular}{cccccccc}   
		 	 \toprule    \multicolumn{8}{c}{True value of ATT: 9.13. Semiparametric Efficiency Bound: 796.8} \\ 
     \midrule          & \multicolumn{7}{c}{Two-way Fixed Effect Estimators} \\
     \cmidrule{2-8}  \noalign{\vskip 1mm}         & Spec. & Avg. Bias & Med. Bias & RMSE  & Asy. Var. & Cover. & CIL \\
     \cmidrule{2-8}    $\tauhat_{fe}$ & Linear & -10.649 & -10.672 & 11.106 & 9907.607 & 0.087 & 12.325 \\   
     $\tauhat_{fe}$ & Saturated & -10.563 & -10.617 & 10.946 & 7924.684 & 0.048 & 11.026 \\  
    \cmidrule{2-8}  \noalign{\vskip 1mm} \multicolumn{7}{c}{Nonparametric Doubly Robust DiD Estimators for the ATT} \\
     \cmidrule{2-8}          & CV Crit. & Avg. Bias & Med. Bias & RMSE  & Asy. Var. & Cover. & CIL \\
     \cmidrule{2-8}    $\tauhat_{dr}$ & ML    & -0.007 & -0.020 & 1.323 & 1721.037 & 0.946 & 5.133 \\ 
     $\tauhat_{dr}$ & LS    & -0.010 & -0.027 & 1.328 & 1732.416 & 0.946 & 5.139  \\ 
     $\tauhat_{sz}$ & ML    & -0.015 & -0.024 & 0.958 & 926.689 & 0.953 & 3.771 \\   
     $\tauhat_{sz}$ & LS    & -0.016 & -0.024 & 0.958 & 926.821 & 0.953 & 3.771  \\  
    \cmidrule{2-8}  \noalign{\vskip 1mm} \multicolumn{8}{c}{Hausman-type test} \\    
   \midrule    & CV Crit. & \multicolumn{1}{l}{Avg. Test Stats.} & \multicolumn{1}{l}{Emp. Size (0.10)} & \multicolumn{1}{l}{Emp. Size (0.05)} & \multicolumn{1}{l}{Emp. Size (0.01)} & & \\ 
   \midrule   & ML    & 1.045 & 0.108 & 0.055 & 0.009   & & \\ 
   &    LS    & 1.045 & 0.107 & 0.056 & 0.009  & & \\    \bottomrule    
		 \end{tabular}
			\begin{tablenotes}[para,flushleft]
				\footnotesize{
					Note: Simulations based on 5,000 Monte Carlo experiments. $ \tauhat_{fe} $ the TWFE regression estimator, $ \tauhat_{dr} $ is our proposed nonparametric DR DiD estimator \eqref{eq.tau.dr.est}, and $ \tauhat_{sz} $ is the nonparametric DR DiD estimator \eqref{eq.tau.sz.est} based on \citet{Sant2020doubly}. For TWFE regression, we use a linear specification, ``Linear'', and a saturated specification, ``Saturated''. For DR DiD estimators, the PS and the OR models are estimated using a local linear logistic and a local linear least squares regression, respectively.  Bandwidth for the PS function is selected with the log-likelihood criterion, ``ML'', and the least squares criterion, ``LS'', respectively.  Lastly, ``Spec.'', ``CV Crit.'', ``Avg. Bias'', ``Med. Bias'', ``RMSE'', ``Asy. Var.'', ``Cover.'', and ``CIL'', stand for the specification, cross-validation criterion,  average simulated bias, median simulated bias, simulated root-mean-squared errors, average of the plug-in estimator for the asymptotic variance, 95\% coverage probability, and 95\% confidence interval length, respectively.  The Hausman-type test statistic is calculated based on \eqref{eq.test.stat}.  Columns ``Avg. Test Stats.'', and ``Emp. Size $ (\alpha) $'' stand for the average test statistic, and empirical size of the test with a nominal size $ \alpha $, respectively. See the main text for further details.}
			\end{tablenotes}
		\end{threeparttable}
\end{adjustbox}\end{table}

In contrast to the results presented in Table \ref{tab:sim.tab.1}, both $ \tauhat_{dr} $ and $ \tauhat_{sz} $ exhibit minimal bias, and their confidence intervals achieve nominal coverage. Their performance is consistently good across different bandwidth selection methods. The TWFE estimators, however, continue to show substantial bias and achieve nearly negligible coverage, despite having much wider confidence intervals compared to the DR DiD estimators.  This occurs because the true treatment effects are heterogeneous, but TWFE specifications do not account for that (i.e., the models are misspecified). In terms of efficiency, the asymptotic variance of $ \tauhat_{sz} $ is reasonably close to the semiparametric efficiency bound.  The asymptotic variance of $ \tauhat_{dr} $ is,  on average, 2.2 times larger than the semiparametric efficiency bound (that imposes no-compositional changes), which is still significantly lower than that of the TWFE estimators. Given that Assumption \ref{ass::stationarity} holds for this DGP, the null hypothesis is true. The empirical rejection frequency of our Hausman-type test is nearly identical to its nominal value, highlighting the desirable properties of this testing procedure.

\subsection{Test power} \label{sec:loc.power}

\begin{figure}[h] 
  \centering 
  \caption{Power curves of the Hausman-type test. Sample Size: $n = 1,000$.}
  \label{fig:power}
  \begin{adjustbox}{width=0.8\textwidth, max height=0.8\textheight, keepaspectratio}
    \begin{threeparttable}
      \centering
      \includegraphics[width=\textwidth]{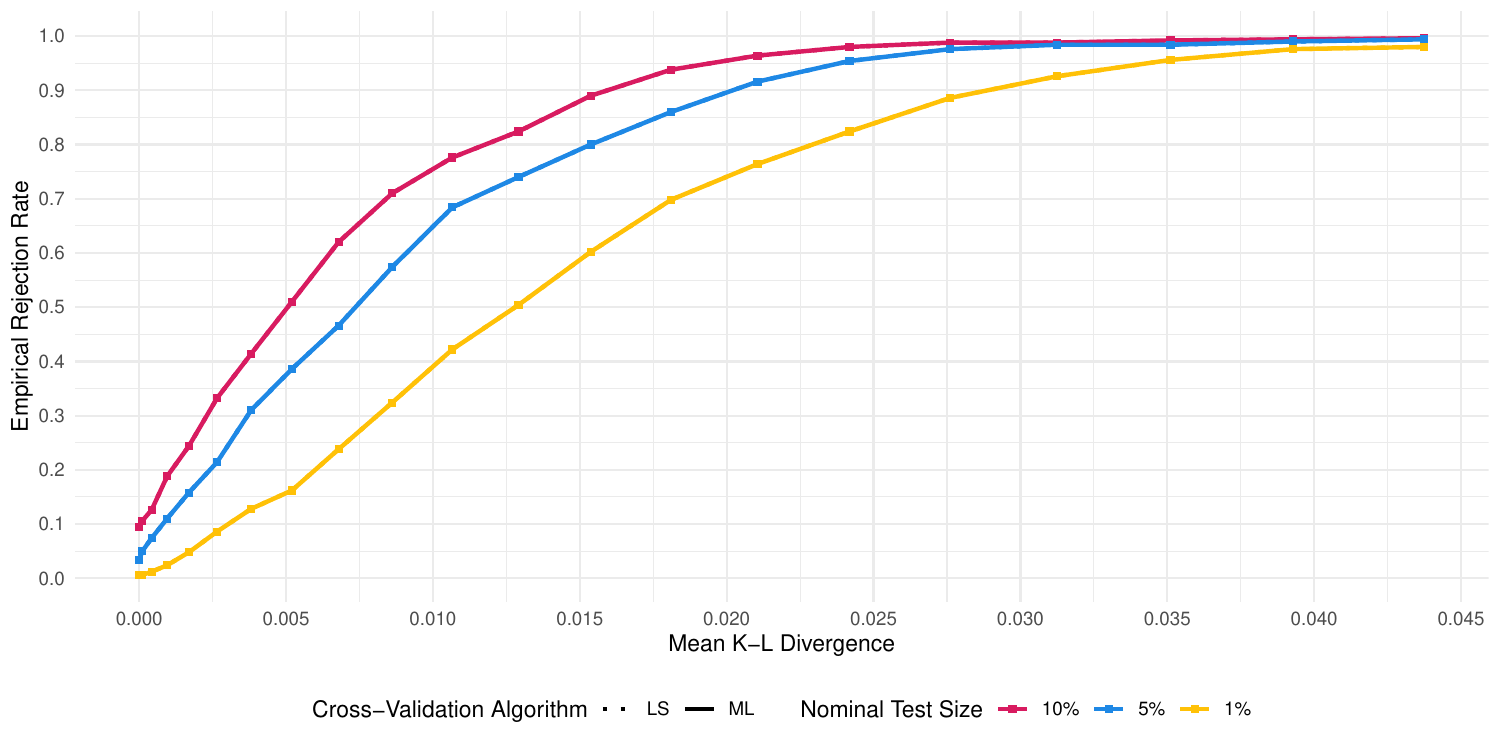}

      \begin{tablenotes}[para,flushleft]
        \footnotesize{
          Notes: Simulations are based on 500 Monte Carlo experiments. The bandwidth for the PS function is selected using LOOCV with two criteria: the log-likelihood criterion (``ML'') and the least-squares criterion (``LS''). The ``mean K-L divergence'' refers to the population mean of the conditional Kullback-Leibler divergence between the propensity scores under the local alternatives and the null hypothesis.
        }
      \end{tablenotes}
    \end{threeparttable}
  \end{adjustbox}
\end{figure}

In this subsection, we examine the power properties of the proposed Hausman-type test. Specifically, we consider a sequence of deviations from the null hypothesis, $\{H_{1,j} \}_{ j = 1}^{20}  $, where the PS functions gradually transition from those in Simulation 1 towards those in Simulation 2. The propensity score functions under these alternatives are generated as follows:
\begin{equation}
    p^{\delta^{alt}_j}(d,t, X) = (1-\delta^{alt}_j) p^{s1} (d,t, X) + \delta^{alt}_j p^{s2} (d,t, X),
\end{equation}
where $\delta^{alt}_j = 0.05 * j $, for $j = 1,...,20$. All other aspects of the DGP remain the same as in Simulation 2. The deviation of the sequence of alternative PS functions can be quantified using the mean K-L divergence between $p^{\delta^{alt}}$ and $ p^{s2} $, 
\begin{equation}
    D_{\mathrm{KL}}(p^{\delta^{alt}} ||~ p^{s2}) = \sE\bracks{\sum_{\dtall}p^{\delta^{alt}}(d,t, X) \log\parens{\dfrac{p^{\delta^{alt}}(d,t, X)}{p^{s2}(d,t, X)}}}.
\end{equation}

Figure \ref{fig:power} presents the power curves for the sequence of local alternatives. It shows that, even when the mean K-L divergence is as small as 0.015, the empirical rejection frequency of the 5\% test is 0.79 (1\%: 0.59, 10\%: 0.88). Additionally, the empirical power does not vary based on the cross-validation criterion used for the PS function: the curves are on top of each other, making them virtually indistinguishable. This demonstrates that the proposed test exhibits robust power properties, even under small deviations from the stationarity assumption.

\subsection{Bandwidth choices} \label{sec:bw.sim}

In this subsection, we present Monte Carlo simulation results comparing three different bandwidth selection methods: LOOCV (described in Section \ref{sec:bw}), RCV (detailed in Appendix \ref{sec:rcv}), and the plug-in estimator (described in Appendix \ref{sec:plugin}).

Table \ref{tab:sim.tab.3} shows that the plug-in bandwidth estimator, which is based on a frequency approach for discrete covariates, exhibits significant bias, elevated variance, and lower empirical test power compared to the other two methods. This is likely due to data sparsity in each stratum when multiple discrete covariates are present.

In contrast, LOOCV and RCV show no noticeable difference in bias across the three DGPs, yet LOOCV demonstrates higher precision, with lower RMSE and shorter confidence intervals. In the non-stationary setting with $\delta^{alt} = 0$, LOOCV reduces RMSE by as much as 10\% and length of 95\% confidence interval by 4.3\% relative to RCV. In terms of hypothesis testing, LOOCV attains an empirical size closer to the nominal level under the null (i.e., $\delta^{alt} = 1$), and achieves higher empirical power as the DGP approaches the non-stationary scenario with $\delta^{alt} = 0$.

 \begin{table}[!ht]
	\caption{Monte Carlo results comparing bandwidth selection methods. Sample size: $n = 1,000$.}%
	\label{tab:sim.tab.3}%
	\centering
	\begin{adjustbox}{ width=0.95\linewidth, max totalheight=0.8\textheight}
		\begin{threeparttable}
            \begin{tabular}{cccccccccc}
            \toprule
            \multicolumn{10}{c}{Non-parametric DR DiD Estimators and Hausman-type Test} \\
            \midrule
                  & \multicolumn{9}{c}{ Non-Stationary Covariate Distribution with $\delta^{alt} = 0$}       \\
        \cmidrule{2-10}          & Avg. Bias  & Med. Bias & RMSE  & Cover.  & CIL   & Avg. Test Stats. & Emp. Pow. (0.10)  & Emp. Pow. (0.05) & Emp. Pow. (0.01) \\
            LOOCV & 0.01  & -0.021 & 1.312 & 0.968 & 5.306 & 21.309 & 1     & 0.999 & 0.98 \\
            RCV   & -0.008 & -0.028 & 1.459 & 0.961 & 5.543 & 21.228 & 0.994 & 0.991 & 0.971 \\
            Plug-in & -2.764 & -2.684 & 13.122 & 0.939 & 49.689 & 1.85  & 0.238 & 0.155 & 0.054 \\ 
             \cmidrule{2-10}  
                  & \multicolumn{9}{c}{Non-Stationary Covariate Distribution with $\delta^{alt} = 0.5$}  \\
        \cmidrule{2-10}          & Avg. Bias  & Med. Bias & RMSE  & Cover.  & CIL   & Avg. Test Stats. & Emp. Pow. (0.10)  & Emp. Pow. (0.05) & Emp. Pow. (0.01) \\
            LOOCV & 0.014 & 0.01  & 1.363 & 0.953 & 5.258 & 6.211 & 0.742 & 0.627 & 0.378 \\
            RCV   & -0.029 & -0.08 & 1.434 & 0.968 & 5.423 & 6.497 & 0.745 & 0.643 & 0.44 \\
            Plug-in & -1.49 & -1.882 & 12.52 & 0.953 & 48.172 & 1.364 & 0.167 & 0.085 & 0.025 \\
                \cmidrule{2-10}     & \multicolumn{9}{c}{Stationary Covariate Distribution with $\delta^{alt} = 1$} \\
        \cmidrule{2-10}          & Avg. Bias  & Med. Bias & RMSE  & Cover.  & CIL   & Avg. Test Stats. & Emp. Size (0.10)  & Emp. Size (0.05) & Emp. Size (0.01) \\
            LOOCV & 0.021 & -0.027 & 1.268 & 0.958 & 5.131 & 1.027 & 0.104 & 0.052 & 0.004 \\
            RCV   & -0.053 & -0.109 & 1.282 & 0.96  & 5.229 & 1.06  & 0.109 & 0.055 & 0.009 \\
            Plug-in & -0.392 & -0.424 & 12.091 & 0.95  & 46.657 & 1.212 & 0.136 & 0.076 & 0.019 \\   \bottomrule    
            \end{tabular}%
			\begin{tablenotes}[para,flushleft]
				\footnotesize{
					Note: Simulations based on 1,000 Monte Carlo experiments based on the three DGPs in Section~\ref{sec:loc.power}, with $\delta^{alt} = 0, 0.5, 1$.  Columns ``Avg. Bias'', ``Med. Bias'', ``RMSE'',  ``Cover.'', and ``CIL'' refer to the average simulated bias, median simulated bias, simulated root-mean-squared errors, 95\% coverage probability, and 95\% confidence interval length for our proposed nonparametric DR DiD estimator, $ \tauhat_{dr} $, as defined in \eqref{eq.tau.dr.est}.  Bandwidth selection is performed via leave-one-out cross-validation (``LOOCV''), rescaled cross-validation (``RCV''), or a plug-in estimator (``Plug-in''), with the first two methods using the log-likelihood criterion in \eqref{eq.cv.ml}. The Hausman-type test statistics are computed according to \eqref{eq.test.stat}.  Columns ``Avg. Test Stats.'', ``Emp. Pow. $ (\alpha) $'', and ``Emp. Size $ (\alpha) $'' denote the average test statistic, empirical power, and empirical size of the test with a nominal size $ \alpha $, respectively. }
			\end{tablenotes}
		\end{threeparttable}
\end{adjustbox}\end{table}

\section{Empirical illustration: the effect of tariff reduction on corruption}\label{sec::application}

In this section, we revisit a study from \citet{sequeira2016corruption} on the effect of import tariff liberalization on corruption patterns.  Prior to the phaseout of high tariffs between South Africa and Mozambique, bribery payment was pervasive, often used to dodge tariff taxes.  According to \citet{sequeira2014corruption},  bribery payments can be found in approximately 80\% of all shipment records in a random sample of tracked shipments before a tariff rate reduction in 2008. 

This tariff change is the result of a long-standing trade agreement between South Africa and Mozambique.  The agreement, the Southern African Development Community Trade Protocol, was signed in 1996. The protocol established a timeline for import tariff reductions between 2001 and 2015.  The most significant reduction occurred in 2008, with the average nominal rate decreasing by 5\%.    The effect of such a tariff liberalization scheme is considerable, as both the likelihood and the amount of bribe payments experienced a significant decline following the phaseout. 


To investigate the causal relationship between tariff rate reduction and changes in bribery patterns,  \cite{sequeira2016corruption} leverages a quasi-experimental variation induced by trade protocol: Not all products were subject to the change in tariff rate during the analysis period, enabling products unaffected by the tariff changes to serve as a control group. It is thus possible to utilize the DiD design to analyze how tariff rate changes affect bribe patterns along trade routes.

  \citet{sequeira2016corruption} collects data on the bribe payment along the trade routes between the two countries from 2007 to 2013.  This data set has a repeated cross-section structure.  \citet{sequeira2016corruption} mainly considers the following two TWFE regressions:
  \begin{align*}
  \text{(Linear) }  \quad & y_{it} =   \gamma_{1} TCC{i} \times Post + \mu Post + \gamma_{2} TCC{i} + \beta_{2} BT_{i} + \Gamma_{i} + p_{i} + w_{t} + \delta_{i} + \epsilon_{it},\\
 \text{(Interactive) }  \quad &y_{it} =   \gamma_{1} TCC{i} \times Post + \mu Post + \gamma_{2} TCC{i} + \beta_{2} BT_{i} + \Gamma_{i} +  \Gamma_{i} \times Post \\
&\quad \quad \quad + p_{i} + w_{t} + \delta_{i} + \epsilon_{it},
  \end{align*}
  where $TCC_i$ and $BT_i$ denote Tariff Change Category and Baseline Tariff, respectively,  and $y_{it} $ is one of the measurements of bribery payments for shipment $ i $ in period $ t $.  $ TCC $ is the treatment indicator, which takes value one if the product shipped experienced a tariff reduction in 2008, and zero otherwise. The post-treatment period indicator, $Post$, is equal to one for the years following 2008. $ BT $ refers to the tariff rates before 2008. A vector of covariates, $ \Gamma $,  industry,  year,  and clearing agent fixed effects, $ p, \omega, \delta $, are also included in the regressions.   The interactive specification differs from the linear one by an interaction of $ Post  $ and the covariates, $ \Gamma$.
  
  \citet{sequeira2016corruption} focuses on interpreting $ \gamma_{1} $ in both specifications as an estimate of the ATT. However, this interpretation might not be valid when treatment effects are heterogeneous \citep{Meyer1995a, Abadie2005}. Our proposed DR DiD estimator, $ \tauhat_{dr} $, and the one based on \citet{Sant2020doubly}, $\tauhat_{sz}  $, could be better suited for the task of identifying and consistently estimating the ATT in the present context. In what follows, we estimate the ATT using our proposed DR DiD estimator and compare the results to those obtained by \citet{sequeira2016corruption}.   
  
  To achieve this, we first estimate the PS and OR functions based on local linear logistic regression and local linear OLS, respectively.  Following  \cite{sequeira2016corruption}, we consider four different  outcome measures:  a binary variable denoting if a bribe is paid, the logarithmic form, $ log(x + 1) $,  of the amount of bribe payment,  the logarithmic form of the amount of bribe paid as a share of the value of the shipment, and as a share of the weight of the shipment, respectively. Across all four specifications, we include the following common covariates: baseline tariff rate, dummy variables for whether the shipper is a large firm, whether the product is perishable, differentiated, an agricultural good, whether the shipments are pre-inspected at origin, monitored, and originates from South Africa. Additionally, we include the day of arrival during the week and the terminal where the cargo was cleared. Our procedures allow for these covariate-specific trends, so the CPT Assumption \ref{ass::ident}(i) holds only after accounting for these observed characteristics. To avoid weak-overlap problems, we truncate PS estimates below 0.01.

  Table \ref{tab:appl.tab.1} summarizes our results. For each estimator, we report both the unclustered standard errors based on asymptotic approximation (in parentheses) and the cluster-robust standard errors based on the bootstrap procedure in Algorithm \ref{alg.boot.se} (in brackets), where we cluster at the four-digit HS code level as in \cite{sequeira2016corruption}. Likewise, we conduct two sets of Hausman-type tests -- one using unclustered influence functions based on \eqref{eq.test.stat} and the other that accounts for clustering using a bootstrap procedure given in Algorithm \ref{alg.boot.test}.

\begin{table}[!ht]
	\caption{Difference-in-Differences estimation results for \citet{sequeira2016corruption} }%
	\label{tab:appl.tab.1}%
	\centering
	%
	\par
	\begin{adjustbox}{ max width=0.9\linewidth, max totalheight=1\textheight, keepaspectratio}
		\begin{threeparttable}
			 \begin{tabular}{lcccc}   
                \toprule    Estimator/Outcome & \multicolumn{1}{l}{Prob(bribe)} & \multicolumn{1}{l}{Log(1 + bribe)} & \multicolumn{1}{l}{Log(1 + bribe/shpt.val.)} & \multicolumn{1}{l}{Log(1 + bribe/shpt.tonn.)} \\ \hline
                \noalign{\vskip 2mm}TWFE - Linear Spec. & -0.429 & -3.748 & -0.011 & -1.914 \\     
                & (0.083) & (0.724) & (0.003) & (0.341) \\ 
                & [0.131] & [1.064] & [0.003] & [0.496] \\
                \noalign{\vskip 2mm}TWFE - Interactive Spec.  & -0.296 & -2.928 & -0.010 & -1.597 \\
                & (0.082) & (0.746) & (0.004) & (0.402) \\  
                & [0.124] & [0.917] & [0.004] & [0.457] \\  
        	 \noalign{\vskip 1mm}DR DiD $\tauhat_{sz}$ &  -0.275 & -2.542 & -0.014 & -0.918 \\      
               (no-compositional changes)  &  (0.067) & (0.636) & (0.005) & (0.451) \\       
                  & [0.096] & [0.773] & [0.006] & [0.492] \\ 
                \noalign{\vskip 2mm} DR DiD  $\tauhat_{dr}$  &  -0.307 & -2.888 & -0.027 & -1.131  \\   
                (robust to compositional changes)  &  (0.084) & (0.798) & (0.010) & (0.602) \\ 
                  & [0.109] & [0.915] & [0.014] & [0.635] \\ \hline 
                   \noalign{\vskip 2mm} \multicolumn{2}{l}{\textbf{Hausman-tests for no-compositional changes}} & & & \\ 
                  \noalign{\vskip 2mm} Unclustered $p$-value & 0.270  & 0.199 & 0.084 & 0.601  \\  
                \noalign{\vskip 2mm} Clustered $p$-value  & 0.338 & 0.238 & 0.175 & 0.643  \\
                  \bottomrule    
			    \end{tabular}%
			\begin{tablenotes}[para,flushleft]
				\footnotesize{
					Notes:  Same data used by \citet{sequeira2016corruption}. The results represent the estimated ATT of tariff rate reduction on bribery payment behavior.  Columns 2 through 5 denote estimates for dependent variables representing whether a bribe is paid, the logarithmic form, $ log(x + 1) $,  of the amount of bribe paid,  the logarithmic form of the amount of bribe paid as a share of the value of the shipment, and as a share of the weight of the shipment, respectively.  We compare four different DiD estimators for the ATT: 1. the two-way fixed effect estimator based on specifications in  Column (1) of Tables 8-11 in \citet{sequeira2016corruption}; 2. the two-way fixed effect estimator based on Column (2) from Tables 8-11 in \citet{sequeira2016corruption}; 3. DR DiD estimator based on \eqref{eq.tau.sz.est}, and 4. DR DiD estimator based on \eqref{eq.tau.dr.est}.  The same set of covariates is used for the last two estimators. See the main text for further details on the covariates.  Continuous variables are re-scaled between 0 and 1, and then added in with binary variables. For DR DiD estimators, the PS and the OR models are estimated nonparametrically, using a local linear least squares and a local linear logistic regression, respectively.  Bandwidth for the local linear logistic regression is selected with the log-likelihood criterion. Numbers in the parentheses are unclustered standard errors based on asymptotic approximation.  Numbers in brackets refer to standard errors clustered at the level of four-digit HS code. Cluster-robust standard errors are calculated following Algorithm \ref{alg.boot.se} with 9999 bootstrap draws. Hausman-tests are calculated based on \eqref{eq.test.stat}. The clustered $p$-values are calculated following the bootstrap procedure in Algorithm \ref{alg.boot.test} with 9999 bootstrap draws. To avoid weak-overlap problems, we truncate PS estimates below 0.01.}
			\end{tablenotes}
		\end{threeparttable}
\end{adjustbox}\end{table}

We first observe that the point estimates are negative for all measures of bribery payment, consistent with the findings of \citet{sequeira2016corruption}. The results based on the two DR DiD methods are generally close to the TWFE estimates with the interactive specification. For instance, we find that a tariff reduction reduces the probability of paying a bribe by 28 to 43 percentage points, depending on the specific estimator used. The result is statistically and economically significant at the usual levels. Tariff reduction also seems to lead to a decrease in bribery.\footnote{Some of local linear OR estimates were a bit sensitive to bandwidth choice. This is arguably due to the limited number of observations within certain strata. To improve the stability of cross-validation, we impose a common bandwidth across all four treatment groups for each type of covariates.} The magnitude of the causal effects based on the weighted results, on the other hand, is more mixed.\footnote{We avoid attaching a precise interpretation of these log transformations due to the issues raised by \citet{Chen2023}.}
Results based on the TWFE and DR DiD with no-compositional changes estimators suggest that tariff reduction leads to a statistically significant reduction in the average log of the ratio between bribery payment and shipment values of similar magnitude, while our proposed DR DiD estimator that is robust to compositional changes suggests a twice-as-large effect. When the log of the ratio between bribery payment and tonnage is considered, both nonparametric DR DiD estimators report large yet insignificant (at 95\% level) ATT estimates. The results of the Hausman-type test displayed at the bottom of Table \ref{tab:appl.tab.1} suggest that we lack statistical evidence against the assumption of no-compositional changes, especially when one clusters the standard errors. 

In sum, our results support the conclusion of \citet{sequeira2016corruption} that tariff liberalization decreases corruption. Our DR DiD estimates suggest the size of the effects is approximately the same as that of the original paper, indicating that ruling out treatment effect heterogeneity and compositional changes are not of primary concern in this particular application. 

\section{Extensions}\label{sec::extensions}
We conclude the paper by considering two extensions of our main results: the use of cross-fitted first-step estimators,  and the analysis of setups with rotating panel data structures where some units are observed in both pre- and post-treatment periods.

\subsection{Cross-fitted first-step estimators}  \label{sec:crossfit}

We describe a cross-fitting procedure for generic first-step estimators. Let \( J \) be a fixed positive integer such that \( J < N \), and assume for simplicity that \( n_J = n / J \) is an integer. Randomly split the dataset into \( J \) equal groups (folds) of size \( n_J \). Denote the set of indices for the \( j \)-th group by \( \Ij \), and let \(\Imj = \{1, \dots, n\} \setminus \Ij \) represent the indices for all observations except those in the \( j \)-th group. For each \( j \), construct the first-step estimators \(\parens{ \phat_j, \{\mhat_{d,t, j}\}_{\dtnpt} }  \) using data from all folds except the \( j \)-th fold, i.e., \(\{W_i\}_{i \in \Imj }\).

The cross-fitted doubly robust estimator, \( \tauhat_{dr, J}^{cf} \), is then defined as:
\begin{align}
\tauhat_{dr, J}^{cf} 
&=  \frac{1}{n}\sumj \sumIj \cbracks{\what_{1, 1}(D_i, T_i)\tauhat_j(Y_i, X_i) +  \sum_{\dtnpt}   (-1)^{(d + t)}  \what_{d,t, j}(D_i, T_i, X_i)  (Y_i - \mhat_{d,t, j}(X_i))},\label{eq.tau.dr.cf}
\end{align}
where $ \tauhat_j(Y, X) =  Y + \sum_{\dtnpt} (-1)^{d+t} \mhat_{d,t, j}(X) $ and  $\what_{d,t, j}(X) = \left. \frac{I_{d,t}\phat_j(1,1,X)}{\phat_j(d,t,X)} \middle / \parens{J^{-1}\sumj\sEj\bracks{\frac{I_{d,t} \phat_j(1,1,X)}{\phat_j(d,t,X)}} } \right. $. Here,  $ \sEj[\cdot]  $ represents the sample average over the observations in $j$-th fold. 

Let   $(\rseq)_{n\geq 1}, (\sseq)_{n\geq 1}$, and $ (\varepsilon'_n)_{n\geq 1} $, be sequences of positive constants approaching 0. We make the following low-level assumptions regarding the cross-fitted estimators.

\begin{asm}[Cross-fitted nuisance estimators] \label{ass:cf.nuisance} \phantom{a}
\begin{enumerate}
\item For any $ j \in \Jset$, the nuisance estimators $ (\phat_j, \{\mhat_{d,t,j}\}_{\dtnpt})$ constructed using $ \dtamj $ belong to the realization set $\Jcal_{n} \equiv \Jcal_{n}^p \times \parens{\Jcal_{n}^{m}}^3 $ with probability at least $ 1- \varepsilon'_n$. The sets $ \Jcal_{n}^p $ and $  \Jcal_{n}^{m}$  include the true nuisance functions and satisfy the following constraints:
\begin{enumerate}[label=$(\roman*)$]
    \item $ \supetap \normL{\ptilde(\cdot, \cdot, \cdot) - p(\cdot, \cdot, \cdot)} \leq \rseq $.
    \item $ \supetam \normL{\mtilde_{d,t}(\cdot) - m_{d,t}(\cdot)} \leq \sseq $, for $\dtnpt$.
    \item $ \rseq \cdot \sseq = \oh(n^{-1/2})$ and $ \rseqo \cdot \sseq = \oh(n^{-1/2})$, for $\dtnpt$.
    \item  $ 0 < \inf_{x \in \Xcal}  \ptilde(d,t, x)  <  \sup_{x \in \Xcal}  \ptilde(d,t, x) <1  $, for $ \dtall $ and $ \ptilde \in \Jcal_{n}^p $.
\end{enumerate}
 \item  $\var{Y|D = 1, T = 1} < \infty $ and $ \sup_{x\in\Xcal}\var{Y|D = d, T = t, X = x}< \infty$, for all $ \dtnpt.$
 \end{enumerate}
\end{asm}

\begin{lem}[Doubly robust error rate with cross-fitted estimators] \label{lem:cf.error.rate} \phantom{Suppose that}
Suppose that Assumptions \ref{ass::sampling}, \ref{ass::ident}, and \ref{ass:cf.nuisance} are satisfied. Then, 
	\begin{equation} 
		\sqrt{n}\parens{\tauhat_{dr, J}^{cf}   - \tau} =  \dfrac{1}{\sqrt{n}}\sum_{i = 1}^{n} \eta_{\text{eff}}(W_{i})  + \op(1) \convd N(0, \Omega_{dr}).
	\end{equation}
\end{lem}

Due to the cross-fitting procedure, Assumption \ref{ass::genr.conv.rate}.2 simplifies to Assumption \ref{ass:cf.nuisance}.1. This assumption now requires that the cross-fitted nuisance estimators converge to their true values in mean square. However, similar to the generic case, Lemma \ref{lem:cf.error.rate} still imposes the requirement that the product of first-stage approximation errors converge at a rate faster than $\Opnsq$. 

Compared to verifying convergence rates for a generic estimator, establishing the mean convergence rate in this specific context is relatively straightforward. Established results provide $L_2 $-rate conditions for a wide range of nonparametric and machine learning estimators, including kernel and series estimators, as well as methods like Lasso, ridge regression, random forests, boosted trees, deep neural networks, and their ensembles.  Although not formalized in this paper, Lemma \ref{lem:cf.error.rate} can be extended to accommodate high-dimensional confounders, where the dimensionality of the nuisance functions increases with the sample size. This setting violates the assumptions on the complexity of the nuisance function space, such as the Donsker properties imposed under Assumption \ref{ass::genr.conv.rate}.2. For a comprehensive discussion of these challenges, refer to \citet{Chernozhukov2017} and references therein.

It is important to note that cross-fitting in the current setup can be challenging when the data size is limited. Since the data for each of the four treatment groups must be split into $ J $ folds, larger values of $ J $ result in fewer observations per fold, thereby increasing estimation error. Another important caveat is that the semiparametric efficiency results may not hold in high-dimensional covariate spaces, a topic we do not cover in this paper. For a detailed discussion of this limitation, see, for example, \citet{jankova2018semiparametric}.

\subsection{Overlapping cross section}\label{sec:ocs}

While the primary focus of this paper is on repeated cross-sectional data where units are observed exclusively in either the pre-treatment or post-treatment period, as per Assumption \ref{ass::sampling}, we recognize that some practical settings involve overlapping cross-sections. For example, surveys like the Current Population Survey (CPS) and the Consumer Expenditure Survey (CEX) employ rotating panel designs. In these surveys, a fraction of the respondents contribute to longitudinal data, appearing in both the pre-and post-treatment periods, while the remaining respondents are observed in only one of the two periods.

In the CPS, households are surveyed for four consecutive months, excluded for the following eight months, and then surveyed again for four more months. Similarly, the CEX follows a rotating panel design where housing units are interviewed once per quarter for four consecutive quarters before being replaced. These designs result in datasets with a mix of panel data and repeated cross-sectional data, where some units overlap across periods, while others are unique to specific periods.

Compared to our assumed sampling process, this structure introduces an additional layer of complexity due to the mixture of unit types. Specifically, let $R = 1$ indicate a panel subject (a unit with data from both pre- and post-treatment periods). The observed data is now  $W^{oc} \equiv ( RY_{0}, RY_{1}, (1-R)Y, (1-R)T, D, X, R)$, where, consistent with the notation introduced in Section \ref{sec:setup}, 
$Y_0$ and $Y_1$ denote the observed outcomes of panel units in periods $0$ and $1$, respectively, 
and $Y$ denotes the observed outcome for each cross-sectional unit. This observed data is described by the following mixture distribution:
\begin{align}
    &\prob{(1-R)Y\leq (1-r)y, RY_1 \leq  ry_{1}, RY_0 \leq ry_{0}, (1-R)T=(1-r)t, D=d, R = r, X\leq x} =  \nonumber \\ & \qquad r \cdot\prob{R = 1} P_{p}(y_{1}, y_{0}, d, x) + (1-r)\cdot\prob{R = 0} P_{rc}(y, d, t, x), \label{eq.dgp.oc}
\end{align}
where 
\begin{align*}
 &P_{p}(y_{1},  y_{0}, d, x) =  \prob{Y_1 \leq y_{1}, Y_0 \leq y_{0},  D=d, X\leq x | R = 1},  \\
 &P_{rc}(y, d, t, x) = \prob{Y\leq y, D=d, T=t, X\leq x| R = 0}.
\end{align*}

The ATT under this setup is defined as 
\begin{small}
    \begin{align*} 
    \tau^{oc} \equiv \prob{R = 1} \sE[ Y_1(1) - Y_1(0) | D = 1, R = 1] + \prob{R = 0} \sE[ Y_1(1) - Y_1(0) | D = 1, T =1, R = 0].
 \end{align*}
 \end{small}
The following conditions are imposed to identify the ATT.

\begin{asm}[Identification assumptions under overlapping cross sections] \label{ass::ident.oc} \phantom{a} 
\begin{enumerate}
\item The observed data $\{W_i^{oc}\}_{i = 1}^{n}$ consists of $ i.i.d.$ draws from the mixture distribution defined in \eqref{eq.dgp.oc}.

\item For some $\varepsilon >0$, $ \dtnpt$, and for almost every $x \in \mathcal{X}$,
\begin{align*}
(p-i) & \  \sE[Y_{1}(0) - Y_{0}(0)|D=1, R=1,X=x] =  \sE[Y_{1}(0) - Y_{0}(0)|D=0, R=1,X=x].\\
(p-ii) & \  \prob{D = 1 | R= 1} > \varepsilon, \text{ and } \prob{D = 1|R = 1, X = x} \leq 1- \varepsilon. \\
(rc-i)& \  \sE[Y_{1}(0)|D=1,T=1,R=0,X=x] - \sE[Y_{0}(0)|D=1,T=0,R=0,X=x] \\
              & = \sE[Y_{1}(0)|D=0,T=1,R=0,X=x]  -\sE[Y_{0}(0)|D=0,T=0,R=0,X=x].\\
(rc-ii)& \   \sE[Y_{0}(0)|D=1,T=0,R=0,X=x] = \sE[Y_{0}(1)|D=1,T=0,R=0,X=x].\\
(rc-iii) & \  \mathbb{P}\left( D=1,T=1| R = 0\right) >\varepsilon, \text{ and } \prob{D=d,T=t|R = 0, X=x} \geq \varepsilon.
\end{align*}
\end{enumerate}
\end{asm}

This new identification assumption amounts to combining the conditions imposed on the panel units (Assumption 2 in \citet{Sant2020doubly}) with those imposed on the cross-sectional units (Assumption \ref{ass::ident}).

The EIF and semiparametric efficiency bound can be derived using arguments analogous to those in Theorem \ref{thm::seb}. Before presenting our results, we introduce the following quantities:
 \begin{align*}
 & \Delta Y = Y_1 -Y_0, \\
 & p_1(x) = \prob{D = 1| R =1, X = x},\  p_0(d,t,x) = \prob{D = d, T= t| R =0, X = x}, \\
&w_1^{p}(D) = \frac{D}{\sE[D | R = 1]} , \quad  w_0^{p}(D, X) = \left. \dfrac{ (1-D) \cdot p_1(X) }{ 1 - p_1(X)}\middle / \sE\bracks{ \left. \dfrac{ (1-D) \cdot p_1(X) }{1- p_1(X) } \right\vert R= 1 }, \right. \\
& m^p_{d, \Delta}(X)  = \sE[\Delta Y|D = d, R = 1, X],  \quad\tau^{oc}_{p} = \sE[m^p_{1, \Delta}(X) - m^p_{0, \Delta}(X) | D = 1, R = 1], \\
&w^{rc}_{1,1}(D, T)  =  \dfrac{DT }{\sE[DT|R=0]} , \quad w^{rc}_{d, t}(D, T, X)  = \left. \dfrac{I_{d,t} \cdot p_0(1,1, X) }{p_0(d,t, X)}\middle / \sE\bracks{\left. \dfrac{I_{d,t}\cdot p_0(1,1,X) }{p_0(d,t, X)} \right\vert R= 0}, \right.  \\
& m_{d,t}^{rc}(X) = \sE[Y|D =d, T = t, R = 0, X],  \quad   \tau^{oc}_{rc}(Y, X)  = Y + \sum_{\dtnpt}(-1)^{(d+t)} m_{d,t}^{rc}(X), \\
& \tau^{oc}_{rc} = \sE[\tau^{oc}_{rc}(Y, X) | D = 1, T= 1, R= 0].
\end{align*}

In addition, let
\begin{align*}
\eta_{p}(Y_1, Y_0, D, X) = &\left\{ w_1^{p}(D)\parens{ m^p_{1, \Delta}(X)  - m^p_{0, \Delta}(X) - \tau^{oc}_{p}} \right. \\
&\quad \left. + w_1^{p}(D) (\Delta Y - m^p_{1, \Delta}(X)) -    w_0^{p}(D, X) (\Delta Y - m^p_{0, \Delta}(X))\right\}, \\
\eta_{rc}(Y, D, T, X) = &  w^{rc}_{1, 1}(D, T)(\tau^{oc}_{rc}(Y, X) -\tau^{oc}_{rc}) +  \sum_{\dtnpt}   (-1)^{(d + t)}  w^{rc}_{d,t}(D, T, X)  (Y - m^{rc}_{d,t}(X)), \\
\eta_r(R) = (R -  \sE[&R] ) \cdot (\tau^{oc}_{p} - \tau^{oc}_{rc} ),
\end{align*}
and
\begin{align*} 
   V_{p}^{oc} & = \sE\left[ D|R= 1\right]^{-2} \cdot  \sE\left[D (m^p_{1, \Delta}(X) - m^p_{0, \Delta}(X) - \tau^{oc}_{p})^2 \right. \\
    & \quad \quad \quad \quad  \left. + D (\Delta Y - m^p_{1, \Delta}(X))^2 
    + \dfrac{(1-D) \cdot p_1(X)^2}{ (1-p_1(X))^2}(\Delta Y - m^p_{0, \Delta}(X))^2 | R = 1 \right],  \\
   V_{rc}^{oc} & = \sE\left[\left. DT\right\vert R=0\right]^{-2}\cdot \sE\left[ DT(\tau^{oc}_{rc}(Y, X) - \tau^{oc}_{rc})^{2} \right.\\
    &\quad \quad  \quad \quad  \left. \left. +\sum_{\dtnpt}\dfrac{I_{d,t} \cdot p_0(1,1,X)^{2}}{p_0(d,t,X)^{2}} (Y - m^{rc}_{d,t}(X))^{2} \right\vert R = 0\right], \\
   V_{r}^{oc} & =    \var{R} \cdot (\tau^{oc}_{p} - \tau^{oc}_{rc})^2.
\end{align*}

\begin{thm}[Semiparametric efficiency bound under overlapping cross sections]\label{thm::seb.oc}
Suppose Assumption  \ref{ass::ident.oc} holds.
\begin{enumerate}
    \item[(a)]  The EIF for the ATT, $\tau^{oc}$ is given by,
\begin{align*}
\eta_{oc}(W^{oc}) = R \cdot \eta_{p}(Y_1, Y_0, D, X) + (1-R)\cdot\eta_{rc}(Y, D, T, X) + \eta_r(R).
\end{align*}
\end{enumerate}

(b) Furthermore, the semiparametric efficiency bound for the set of all regular estimators of $ \tau^{oc} $ is 
\[ 	\sE[\eta_{oc}(W^{oc})^{2}]  = \sE[R]\cdot V_{p}^{oc} +  \sE[1-R]\cdot V_{rc}^{oc} + V_{r}^{oc}.\] 

\end{thm}

The EIF comprises three parts. The first two correspond, respectively, to the EIF for panel-only data (as derived in Proposition 1(a) of \citet{Sant2020doubly}) and the EIF for cross-sectional–only data (as derived in Theorem \ref{thm::seb} of our paper). The third component directly quantifies the relative effect of assignment to the panel units on average and reflects the efficiency cost associated with combining the two data types.

Interestingly, the semiparametric efficiency bound decomposes into three distinct terms, each reflecting the independent contribution of a different component of the EIF. Notably, when the two data sources share the same true ATT (e.g., when there are no compositional shifts between different sampling cohorts over time), the third term vanishes.

Leveraging this result, we propose the following DR estimand for the ATT:
\begin{align*}
 \tau^{oc}_{dr} = & \sE\left[ R\cdot \left( w_1^{p}(D)\parens{ m^p_{1, \Delta}(X)  - m^p_{0, \Delta}(X)} \right. \right. \\
&\quad\quad\quad\quad\quad \left. + w_1^{p}(D) (\Delta Y - m^p_{1, \Delta}(X)) -    w_0^{p}(D, X) (\Delta Y - m^p_{0, \Delta}(X))\right) \\
&\quad  \left. + (1-R)\cdot  \left( w^{rc}_{1, 1}(D, T)\tau^{rc}(Y, X) +  \sum_{\dtnpt}   (-1)^{(d + t)}  w^{rc}_{d,t}(D, T, X)  (Y - m^{rc}_{d,t}(X)) \right) \right].
\end{align*}

Since $\tau^{oc}_{dr} $ is based on the EIF, it is straightforward to show that plug-in estimators for the ATT based on $\tau^{oc}_{dr}$ inherits the same rate doubly robust property as established in Theorem \ref{thm::asy.normal}. Adapting our paper's arguments in Theorem \ref{thm::asy.normal} to show these results is straightforward. We omit the details for brevity.

\small{
\setlength{\bibsep}{1pt plus 0.3ex}
\putbib
}
\end{bibunit}


\end{document}